\documentclass[cernpreprint,texlive=2011,txfonts,UKenglish,subfigure=true]{latex/atlasdoc}
  \usepackage{latex/atlaspackage}
  \usepackage{latex/atlasphysics}
  \usepackage{latex/atlasbiblatex}
\pdfoutput=1
\usepackage[T1]{fontenc}

\usepackage{epsfig}
\usepackage{epstopdf}
\usepackage{xspace}
\usepackage{rotating}
\usepackage{array}
\usepackage{units}

\usepackage{ifpdf}
\ifpdf
  \usepackage{pdflscape}
\else
  \usepackage{lscape}
\fi

\usepackage{siunitx}

\def\rts {\ensuremath{\sqrt{s}}}

\newcommand{\wma}{\ensuremath{\sin^2{\theta_{\mathrm eff}^{\textrm{~lept}}}}}
\newcommand{\wmanolep}{\ensuremath{\sin^2{\theta_{\mathrm eff}}}}

\newcommand{\wmaqmu}{\ensuremath{\sin^2{\theta_{\mathrm eff}^{\textit{\,q/$\mu$}}}}}
\newcommand{\wmamu}{\ensuremath{\sin^2{\theta_{\mathrm eff}^{\textit{\,$\mu$}}}}}
\newcommand{\wmaq}{\ensuremath{\sin^2{\theta_{\mathrm eff}^{\textit{\,q}}}}}
\newcommand{\cosTh}{\ensuremath{\cos{\theta}}}
\newcommand{\swsq}{\ensuremath{\sin^2{\theta_{\mathrm W}}}}
\newcommand{\sinsqeff}{\ensuremath{\sin^2{\theta_{\mathrm eff}^{\textrm{~lept}}}}}
\newcommand{\cosThSq}{\ensuremath{\cos^{2}{\theta}}}
\newcommand{\cosThCS}{\ensuremath{\cos{\theta_{\mathrm{CS}}^{*}}}}

\newcommand{\dsdcos}{\ensuremath{\frac{\mathrm{d}\sigma}{\mathrm{d}(\cos{\theta})}}}

\newcommand{\Afb}{\ensuremath{A_{\mathrm{FB}} }}
\newcommand{\afb}{\Afb}

\newcommand{\AmuPar}{\ensuremath{A_{\mu}}}
\newcommand{\zgamma}{\ensuremath{Z/\gamma^*}}

\newcommand{\ztau}{\ensuremath{\zgamma\rightarrow\tau\tau}}

\newcommand{\sectionname}{section}
\newcommand{\figname}{figure}

\newcommand{\Figname}{Figure}
\newcommand{\Fignames}{Figures}
\newcommand{\tabname}{table}

\newcommand{\eqname}{eq.}

\newcommand{\wmaMuStat}{0.0009\xspace}
\newcommand{\wmaMuVATLAS}{0.2307\xspace}
\newcommand{\wmaMuSysNoPDFATLAS}{0.0008\xspace}\newcommand{\wmaMuSysPDFATLAS}{0.0009\xspace}
\newcommand{\wmaMuErrorATLAS}{0.0015\xspace}

\newcommand{\AmuParATLAS}{0.153\xspace} \newcommand{\AmuParErrorStatATLAS}{0.007\xspace}
\newcommand{\AmuParErrorSysATLAS}{0.006\xspace}
\newcommand{\AmuParErrorPDFATLAS}{0.007\xspace}
\newcommand{\AmuParErrorTotATLAS}{0.012\xspace}

\newcommand{\fitquality}{22.6/16}

\newcommand{\wmaElStat}{0.0009\xspace}\newcommand{\wmaChiCC}{22.4/16\xspace}

\newcommand{\wmaElVATLAS}{0.2302\xspace}

\newcommand{\wmaElSysNoPDFATLAS}{0.0008\xspace}\newcommand{\wmaElSysPDFATLAS}{0.0010\xspace}
\newcommand{\wmaElErrorATLAS}{0.0016\xspace}

\newcommand{\wmaElStatCF}{0.0007\xspace}\newcommand{\wmaChiCF}{21.9/16\xspace}
\newcommand{\wmaElVCFATLAS}{0.2312\xspace}
\newcommand{\wmaElSysNoPDFCFATLAS}{0.0008\xspace}\newcommand{\wmaElSysPDFCFATLAS}{0.0010\xspace}
\newcommand{\wmaElErrorCFATLAS}{0.0014\xspace}

 \newcommand{\wmaCombStat}{0.0005}

\newcommand{\wmaCombATLAS}{0.2308} 
\newcommand{\wmaCombSysNoPDFATLAS}{0.0006\xspace}\newcommand{\wmaCombSysPDFATLAS}{0.0009\xspace}
\newcommand{\wmaCombErrorATLAS}{0.0012}

\newcommand{\wmaCombEATLAS}{0.2308}
\newcommand{\wmaCombEStat}{0.0006}

\newcommand{\wmaCombESysNoPDFATLAS}{0.0007\xspace}\newcommand{\wmaCombESysPDFATLAS}{0.0010\xspace}
\newcommand{\wmaCombEErrorATLAS}{0.0013}

\AtlasTitle{Measurement of the forward-backward asymmetry of electron and muon pair-production in $pp$ collisions at  $\rts$  = 7 TeV with the ATLAS detector}
\PreprintIdNumber{CERN-PH-EP-2014-259}
\AtlasJournal{JHEP}

\author{The ATLAS Collaboration}

\AtlasAbstract{
This paper presents measurements from the ATLAS experiment of the forward-backward asymmetry in the reaction
$pp\rightarrow Z/\gamma^{*}\rightarrow l^{+}l^{-}$, with $l$ being electrons or
muons, and the extraction of the effective
weak mixing angle.
The results are based on the full set of data collected in 2011 in $pp$ collisions at the LHC at $\rts$ = 7 TeV,
corresponding to an integrated luminosity of 4.8 fb$^{-1}$.
The measured asymmetry values are found to be in agreement with the corresponding Standard Model predictions.
The combination of the muon and electron channels yields a value of the effective weak mixing angle of
$\wma = \wmaCombATLAS  \pm \wmaCombStat  \mathrm{(stat.)} \pm \wmaCombSysNoPDFATLAS \mathrm{(syst.)}
\pm \wmaCombSysPDFATLAS \mathrm{(PDF)}$, where the first uncertainty corresponds to data statistics,
the second to systematic effects and the third to knowledge of the parton density functions. This result
agrees with the current world average from the Particle Data Group fit.
}

\begin{document}
\maketitle
\tableofcontents
\newpage

\section{Introduction}
\label{sec:intro}
The vector and axial-vector couplings in the neutral current annihilation process
$q\bar{q}\rightarrow Z/\gamma^*\rightarrow \ell^{+}\ell^{-}$
lead to a forward-backward asymmetry $\Afb$ in the polar angle distribution of
the final state lepton $\ell^{-}$ with respect to the quark direction in the
rest frame of the dilepton system.
This paper presents measurements of the forward-backward asymmetry in electron and muon pairs
from $\zgamma$~boson decays and the extraction of the weak mixing angle
by the ATLAS experiment.
The results are based on the full set of $pp$ collision data collected in 2011 at the LHC at
a centre-of-mass energy of $\rts$ = 7\,TeV,
corresponding to an integrated luminosity of 4.8~fb$^{-1}$.

The differential cross section for the annihilation process can be written at leading order as

\begin{equation}\label{eq:dsdcos}
\dsdcos =
\frac{4\pi\alpha^2}{3\hat{s}}\left[\frac{3}{8}A(1+\cosThSq)+B\cosTh\right],
\end{equation}

\noindent where $\alpha$ is the fine-structure constant, $\sqrt{\hat{s}}$ is the centre-of-mass energy of the quark and anti-quark, and $\theta$ is
the angle between the lepton and the quark in the rest frame of the dilepton system. The coefficients $A$ and $B$
are functions of $\sqrt{\hat{s}}$ and of the electroweak vector and axial-vector couplings.
In the case that the dilepton system has non-vanishing transverse momentum,
$p_{\textrm{T}}$, the four-momentum of the incoming (anti-)quark is not known, as it is no longer
collinear with the incoming beams.
The impact of this effect on the asymmetry measurement is minimized by choosing a particular
rest frame of the dilepton system, the  Collins--Soper (CS) frame~\cite{CSFrame}, in which the angle between the lepton and
the quark, $\theta^{*}_{\mathrm{CS}}$, is calculated.
The sign of \cosThCS~is defined with respect to the direction of the quark, which is, however, ambiguous
in $pp$ collisions. It is therefore chosen by measuring the longitudinal
boost of the final-state dilepton system in the laboratory frame, and assuming that this
is in the same direction as that of the quark in the initial state. This assumption leads
to a fraction of events with wrongly assigned quark direction, which causes a dilution
of the observed asymmetry. The probability of correct quark direction assignment increases
with the boost of the dilepton system, thus reducing the dilution for dileptons produced at large rapidities.
With this assumption, \cosThCS~ can be written as a function of
the lepton momenta in the laboratory frame,

\begin{equation}\label{eq:costhetastar}
\cosThCS=  \frac{p_{\textrm{z}, \ell \ell}}{|p_{\textrm{z},\ell \ell}|} \frac{2 (p_1^+p_2^--p_1^-p_2^+)}{m_{\ell \ell}\sqrt{m_{\ell \ell}^2+p_{\textrm{T},\ell \ell}^2}}
\end{equation}

\noindent with

\begin{displaymath}\label{eq:costhetastar:P}
 p_{\textit{i}}^\pm=\frac{1}{\sqrt{2}}(E_{\textit{i}}\pm p_{\textrm{z},\textit{i}}),
\end{displaymath}

\noindent where $E$ is the energy and $p_{\textrm{z}}$ the longitudinal momentum of
the lepton ($i=1$) and anti-lepton ($i=2$). The variables $p_{\textrm{z}, \ell \ell}$, $m_{\ell \ell}$,
and $p_{\textrm{T}, \ell \ell}$
denote the longitudinal momentum, invariant mass and transverse momentum of the dilepton system,
respectively.
The first factor in \eqname~\ref{eq:costhetastar} defines the sign of \cosThCS\ according to the longitudinal
direction of flight of the dilepton system, as discussed above.
The events with \mbox{$\cosThCS\geq0$}~are classified as forward (F), while those
having \mbox{$\cosThCS<0$}~are classified as backward (B). The asymmetry \Afb~is then defined as

\begin{equation}\label{eq:afb}
\Afb = \frac{\sigma_{\mathrm{F}}-\sigma_{\mathrm{B}}}{\sigma_{\mathrm{F}}+\sigma_{\mathrm{B}}},
\end{equation}

\noindent where $\sigma_{F}$ and $\sigma_{B}$ are the cross sections for the respective forward and backward configurations.
At leading order, the second term in \eqname~\ref{eq:dsdcos}, $B\cos\theta$, describes the
asymmetry \Afb.
This analysis measures \Afb\  as a function of the invariant mass of the dilepton system.
The results, which are presented in \sectionname~\ref{sec:afbmeas}, include the detector-level values, as well
as the corrections needed to take into account detector effects and dilution.

Several Standard Model parameters can be extracted from the dependence of the \Afb~values on the
invariant dilepton mass. One of these is
the electroweak mixing angle $\swsq$, which is defined at tree level as ${1-m_{W}^2/m_{Z}^2}$.
Depending on the renormalisation scheme, higher-order loop corrections  may modify this relation.
This analysis extracts the effective leptonic weak mixing angle \wma~\cite{PhysRevD.22.971,EWKmeasurementsZ}
at the $m_{Z}$ scale from the detector-level \Afb~values.

The effective weak mixing angle is related to the electroweak vector coupling $\bar{g}_V^f$ via
\[ \bar{g}_V^f = \sqrt{\rho_f} \left( T^3_f - 2Q_f\wmanolep
\right), {\mathrm \,\,\, with \,\, }   \wmanolep =
\kappa_f\sin^2\theta_W,
\]
where the electroweak radiative corrections to the tree-level couplings are absorbed into the fermion-dependent
factors $\kappa_f$ and $\rho_f$, $T^3_f$ is the third component of the weak isospin and
$Q_f$ the electric charge of the fermion $f$.
Using this definition of the
effective weak mixing angle, the coupling retains its tree-level form multiplied by the
additional factor $\sqrt{\rho_f}$.
The relationship between the leptonic and quark \wmanolep\ can be approximated as a flavour-dependent shift in the
leptonic \wmanolep~\cite{EWKmeasurementsZ}.
Although the \wmanolep\ value from $b$-quarks differs the most from the one from leptons, only a few percent of the
events in this analysis come from initial-state $b$-quarks. In particular, the effect of the quark \wmanolep\ on the measured \Afb\
is an order of magnitude smaller than the effect of the leptonic \wmanolep.
This analysis therefore measures the leptonic \wmanolep, denoted by \wma\ in the following.
Its value is extracted from the measured \Afb\ as a function of the invariant mass of the dilepton system by comparing
it to MC predictions produced with varying values of the weak mixing angle. Details are given in
\sectionname~\ref{sec:wma}.

The most precise measurement of \wma\ comes from the combination of results from the
LEP and SLD experiments~\cite{EWKmeasurementsZ}.
Those studies yield an average leptonic $\wma\ = 0.23153 \pm 0.00016$.
The two most precise single measurements are extracted from the
forward-backward asymmetry in $b$-quark final states, $\Afb^{0,b}$, at LEP ($\wma=0.23221 \pm 0.00029$)
and from the leptonic left-right polarization asymmetry, $A_{\mathrm LR}$, at SLD ($\wma=0.23098 \pm 0.00026$).
These two values differ by approximately three standard deviations.
More recently, the CDF~\cite{CDFsin2ThWMu} and D0~\cite{D0sin2ThW} experiments at the Tevatron and the CMS~\cite{CMSNote, CMSOther}
experiment at the LHC have also measured \wma .
The CDF (D0) measurement was performed using $Z \to \mu\mu$ ($Z \to ee$) events from $p\bar{p}$ collisions,
and the CMS measurements were performed using $Z \to \mu \mu $ events from $pp$ collisions.
These results are compared to those from this analysis in \sectionname~\ref{sec:wma}.

The value of \Afb\ at the peak of the \zgamma\ resonance ($m_{\ell \ell}=m_{Z}$), $\Afb^{0,\ell}$, can be written as a function of the asymmetry parameters
$A_\ell$ and $A_{q}$,

\begin{equation}\label{eq:ai}
\Afb^{0,\ell} = \frac{3}{4}A_q A_\ell ,
\end{equation}

\noindent with $\ell$ ($q$) denoting the leptons (quarks) in the final (initial) state.
The parameters $A_\ell$ and $A_{q}$ are directly related to the electroweak vector and axial-vector
couplings, as described in \sectionname~\ref{sec:Amu}.
The most precise measurements of the electron and muon asymmetry parameters
were performed by SLD~\cite{EWKmeasurementsZ}, yielding $A_e = 0.15138 \pm 0.00216$  and $\AmuPar = 0.142 \pm 0.015$.
The precision of the
\AmuPar\ measurement is dominated by the statistical uncertainty, thus making it an interesting parameter to measure
with the large number of $Z \to \mu\mu$ events produced at the LHC. The \AmuPar\ result
from this analysis is presented in \sectionname~\ref{sec:Amu}.
The determination
of \AmuPar\ in the LEP/SLD results is entirely based on asymmetry measurements in
electron and muon final states without any assumptions on the involved $A_f$. In contrast, the
determination of \AmuPar\ presented here uses the Standard Model prediction of $A_q$.

 \section{The ATLAS detector}
The ATLAS detector~\cite{DetPaper} is a general-purpose detector installed at the LHC~\cite{LHC} at CERN.
 The detector subsystem closest to the interaction point, the inner detector (ID), provides
 precise position and momentum measurements of charged particle trajectories. It covers the pseudorapidity range\footnote{ATLAS
 uses a right-handed coordinate system with its origin at the nominal interaction
 point (IP) in the centre of the detector and the $z$-axis along the beam pipe. The $x$-axis
 points from the IP to the centre of the LHC ring, and the $y$-axis points upward. Cylindrical
 coordinates ($r$,$\phi$) are used in the transverse plane, $\phi$ being the azimuthal angle around the
 beam pipe. The pseudorapidity is defined in terms of the polar angle $\theta$ as $\eta= -\ln \tan(\theta/2)$.}
 $|\eta|<2.5$ and provides full azimuthal
 coverage. The ID consists of three subdetectors arranged in a coaxial geometry
 around the beam axis: the silicon pixel detector, the semiconductor microstrip detector
 and the straw-tube transition-radiation tracker. A solenoidal magnet generates a 2\,T
 magnetic field in which the ID is immersed.

 Electromagnetic calorimetry in the region $|\eta|<3.2$ is based on a
 high-granularity, lead/liquid-argon (LAr) sampling technology. Hadronic
 calorimetry uses a scintillator-tile/steel detector in the region $|\eta|<1.7$
 and a copper/LAr detector in the region $1.5<|\eta|<3.2$. The most forward region of the
 detector ($3.1<|\eta|<4.9$) is equipped with a forward calorimeter,
 measuring both the electromagnetic and hadronic energies using copper/LAr
 and tungsten/LAr modules.

 A large muon spectrometer (MS) constitutes the outermost part of the detector.
 It consists of three large air-core superconducting toroidal magnet systems (each with eight coils):
 one barrel providing a field
 of about 0.5\,T and two endcaps each providing a field of about 1\,T. The deflection of the muon
 trajectories in the magnetic field is measured in three layers of
 precision drift tube chambers for $|\eta|<2$. In higher |$\eta$| regions ($2.0<|\eta|<2.7$), two layers of drift tube chambers are used in combination
 with one layer of cathode strip chambers in the innermost endcap wheels of the MS.
 Three layers of resistive plate chambers in the barrel ($|\eta|<1.05$)
 and three layers of thin gap chambers in the endcaps ($1.05<|\eta|<2.4$) provide
 the muon trigger and also measure the muon trajectory in the non-bending plane.

 A three-level trigger system is used to select events in real time. A hardware-based
 Level-1 trigger uses a subset of detector information to reduce the event rate to a design
 value of at most 75 kHz. The rate of accepted events is then reduced to about 300\,Hz by two software-based
 trigger levels, Level-2 and the Event Filter.

\section{Signal and background modelling}
\label{sec:dataMC}

Monte Carlo (MC) simulated event samples used to model signal and background processes
are generated and passed through the ATLAS detector simulation~\cite{SimPaper}, based on the GEANT4 toolkit~\cite{geant}.
Simulated events acquire weights such that the resulting distributions match the ones observed
in the data for the following variables: the average
number of interactions per bunch crossing, the $z$ coordinate of the interaction vertex, the lepton energy/momentum
scale and resolution, and the trigger, identification and reconstruction efficiencies.

The \zgamma\ production is detected by the emission of charged lepton pairs, $ee$ or $\mu\mu$.
The contribution from \ztau\ followed
by $\tau$ decays to electrons or muons is considered as background and subtracted from the signal.
Signal samples are generated with PYTHIAv6.4~\cite{pythia} with the
MSTW2008LO~\cite{mstw} parton distribution functions (PDFs) and a value of
$\wma = 0.232$ for the effective weak mixing angle. Final-state radiation from QED is taken
into account using PHOTOS~\cite{photos}~in the exponentiated mode with multi-photon emission.
For the \wma\ measurement, where sensitivity to PDFs is expected to be significant, additional
PDF sets are also used, including one specifically prepared for this analysis, based on the ATLAS-epWZ12 PDFs~\cite{epWZpdf}.
Details are given in \sectionname~\ref{sec:wma}.
The cross section is calculated at next-to-next-to-leading order (NNLO) in the strong coupling using
PHOZPR~\cite{phozpr} with MSTW2008 NNLO PDFs.
The ratio of this cross section to the leading-order (LO) cross section is
the $m_{\ell \ell}$-dependent $K$-factor, applied to the generated signal for all plots shown in the following.
However, the main observable described here (\Afb) is not affected by this LO-to-NNLO rescaling.
The impact of higher-order corrections in $\alpha_{\mathrm{s}}$ and $\alpha_{\mathrm{em}}$ on \Afb\ and on \wma\ is assessed using the HORACEv3.1~\cite{horace}, MCFMv6.6~\cite{mcfm} and POWHEGv1~\cite{powheg}
generators, as described in \sectionname~\ref{sec:afbmeas}. The POWHEG simulation is combined
with PYTHIA 6.4 for showering and hadronization.

Background contributions containing prompt isolated electron or muon pairs are estimated using
Monte Carlo simulation normalized using the best available cross section prediction at (N)NLO.
The background from \ztau~is also generated using PYTHIAv6.4.
Diboson ($WW,~WZ,$ and $ZZ$) samples are generated with HERWIGv6.510~\cite{herwig2,herwig} using the
MRSTMCal PDFs \cite{MRSTMcal}.
Pair-production of top quarks is generated with MC$@$NLOv4.01~\cite{mcnlo,mcnloMatching} using the CTQE6L1 PDFs
\cite{CTEQ6L1},
combined with HERWIG for showering and hadronization.

The contributions from multi-jet and $W$+jets background events containing non-isolated leptons from heavy-flavour
decays and hadrons misidentified as leptons are estimated using data-driven techniques, as described in \sectionname~\ref{sec:eventselection}. Since
the contribution from $W$+jets is found to be a small fraction of the multi-jet background over the whole invariant mass range,
the term `multi-jet background' is used in the following to denote the sum of these contributions.

\section{Event reconstruction and selection}
\label{sec:recoSel}

The analysis uses $pp$ data collected in 2011, corresponding to an integrated luminosity
of 4.8\,fb$^{-1}$ for the electron channels and 4.6\,fb$^{-1}$ for the muon channel.
All events analysed were acquired under good operating conditions of the ATLAS detector.
Events in the electron channels passed the single electron trigger, with an electron
$E_{\textrm{T}}>$ 20 or 22\,GeV
(depending on the instantaneous luminosity).
Events in the muon channel passed the single muon trigger with a muon $p_{\textrm{T}}$ threshold of 18\,GeV.
The presence of a reconstructed collision vertex with at least three
tracks with $p_{\textrm{T}} > 400$\,MeV is required. For the muon channel, there is an additional requirement that the
longitudinal position of this vertex be within 200\,mm of the nominal interaction point.

\subsection{Electron reconstruction}
\label{sec:elereco}

This analysis uses electrons in two distinct regions of the detector: the central region (\hbox{$|\eta|<2.47$})
where tracking information is available, and the forward region (\hbox{$2.5<|\eta|<4.9$}), where the
electron reconstruction relies only on information from the calorimeter. The inclusion of
electrons in the forward region allows
the reconstruction of events where the $\zgamma$ candidates are emitted at large rapidity,
thus reducing the effect of dilution due to the imperfect knowledge of the direction of the initial state quark.

For both the central and forward electrons, the reconstruction begins with
identifying energy deposits in the calorimeters consistent with electromagnetic showers.
Electron candidates in the central region are matched to a track reconstructed in the ID.
A transverse energy requirement, $E_{\textrm{T}}>$ 25\,GeV, is applied to both the central and forward candidates.
Electron candidates in transition regions between the barrel and endcap calorimeters ($1.37<|\eta|<1.52$)
and between the endcap and forward calorimeters ($3.16<|\eta|<3.35$) are excluded from this analysis.

The central candidates must satisfy either `medium' or `tight' identification criteria, based on shower
shape and
track quality variables~\cite{EurPhysJC72} optimized for the 2011 data~\cite{2011ElPerf}.
Forward electron candidates
must satisfy similar medium quality criteria optimized specifically for forward electrons.
When combining one central and one forward electron there is an additional
requirement that the central electron be isolated. It is implemented by requiring that the
transverse energy
deposition in the calorimeter within
a cone defined by $\Delta R~ =\sqrt{\left( \Delta\phi\right)^2+\left(\Delta\eta\right)^2}=0.2$
around the electron candidate be less than 5\,GeV, excluding the electron candidate itself.
The reconstruction efficiencies in simulated events are corrected to match the measured efficiencies~\cite{2011ElPerf}.
Selected events consist of either two medium candidates in the central region (central--central, referred to as CC) or
one tight central electron candidate and one medium forward electron candidate (central--forward,
referred to as CF). In the CC electron channel, the
electrons are required to have opposite charges. This requirement is not applicable in the CF electron channel, since the forward electron has
no charge information. The effect of charge misidentification is found to be negligible in both the CC and CF electron channels~\cite{2011ElPerf}.

\subsection{Muon reconstruction}
\label{sec:muonreco}

Muons identified as `combined muons' by the reconstruction and identification algorithms~\cite{muonPerformance,MuonPerfRun1} are used
in this analysis. They are reconstructed by associating and combining two independently reconstructed tracks,
one in the ID
and one in the MS. Combined muons are required to have transverse momentum $p_{\textrm{T}}>$ 20\,GeV, and must lie within ${|\eta|<2.4}$.
The ID tracks associated with the muons must satisfy quality requirements
on the number of hits recorded by each subdetector~\cite{muonPerformance}.
To reject muons from cosmic rays, the longitudinal coordinate of
the point of closest approach of the track to the beamline is required to be within 10\,mm of the collision vertex (see \sectionname~\ref{sec:eventselection}).
Rejection of multi-jet background is improved by
requiring the muons to be isolated. The isolation
parameter is the relative momentum isolation, defined as the sum of the $p_{\textrm{T}}$
of all other tracks within a cone of $\Delta R = 0.2$ around the muon
track, divided by the muon $p_{\textrm{T}}$: ${\sum p_{\textrm{T}}^{\mathrm track} / p_{\textrm{T}}^{\mu} < ~\mathrm{0.1}}$.
The kinematic variables of the muons are measured by the ID, in order to minimize the impact of residual misalignments
between the ID and the MS. This choice also reduces the impact of muon bremsstrahlung in the calorimeter on the measurement.
Charge misidentification for muons is very low, with negligible effect on this analysis.

\subsection{Event selection and background estimation}
\label{sec:eventselection}

Events must contain two oppositely charged leptons in the muon and CC electron channels or one central electron
and one forward electron in the CF electron channel.
In the muon and CC electron channels, dilepton pairs with invariant masses up to 1000\,GeV are used.
In the CF electron channel, the \Afb~measurement is performed only for dilepton masses up to 250\,GeV, because
the background dominates at larger masses, leading to sizeable systematic uncertainties.

Contributions of different background sources are estimated using either simulation or data-driven techniques.
For dibosons ($ZZ$, $WZ$, $WW$), $\ztau$  and $\ttbar$, contributions are estimated using simulation. The dominant background
for the muon and CC electron channels, across the whole invariant mass range, is that due to $\ttbar$ events.
Background from \ztau\ production (followed by $\tau\rightarrow \ell \nu$) populates the low end of the dilepton invariant mass distribution.
For the electron channels, the multi-jet background is estimated using
a combination of data-driven techniques. In the CC electron channel, the reverse identification method~\cite{WZproduction}
is used for dilepton invariant masses below 125\,GeV, while the
fake-factor method~\cite{ZprimePaper}, is employed for higher invariant masses. An overlap region is defined between 110
and 200 GeV, where the estimates from both methods are compared and a scale factor for the reverse identification estimate is determined using the
fake-factor result.
Since the CF electron channel only extends to a dilepton invariant mass of 250 GeV, only the reverse identification method is used.
In the muon channel, the multi-jet background is estimated from data in a control region
defined by inverting the muon isolation cut. The numbers of events in the control and signal regions
observed in MC simulation are then used to transfer the data distribution from the control region to
the signal region.

\begin{figure}[p!]
  \begin{center}
    \subfigure[]{      \includegraphics[width=0.444\textwidth,angle=0]{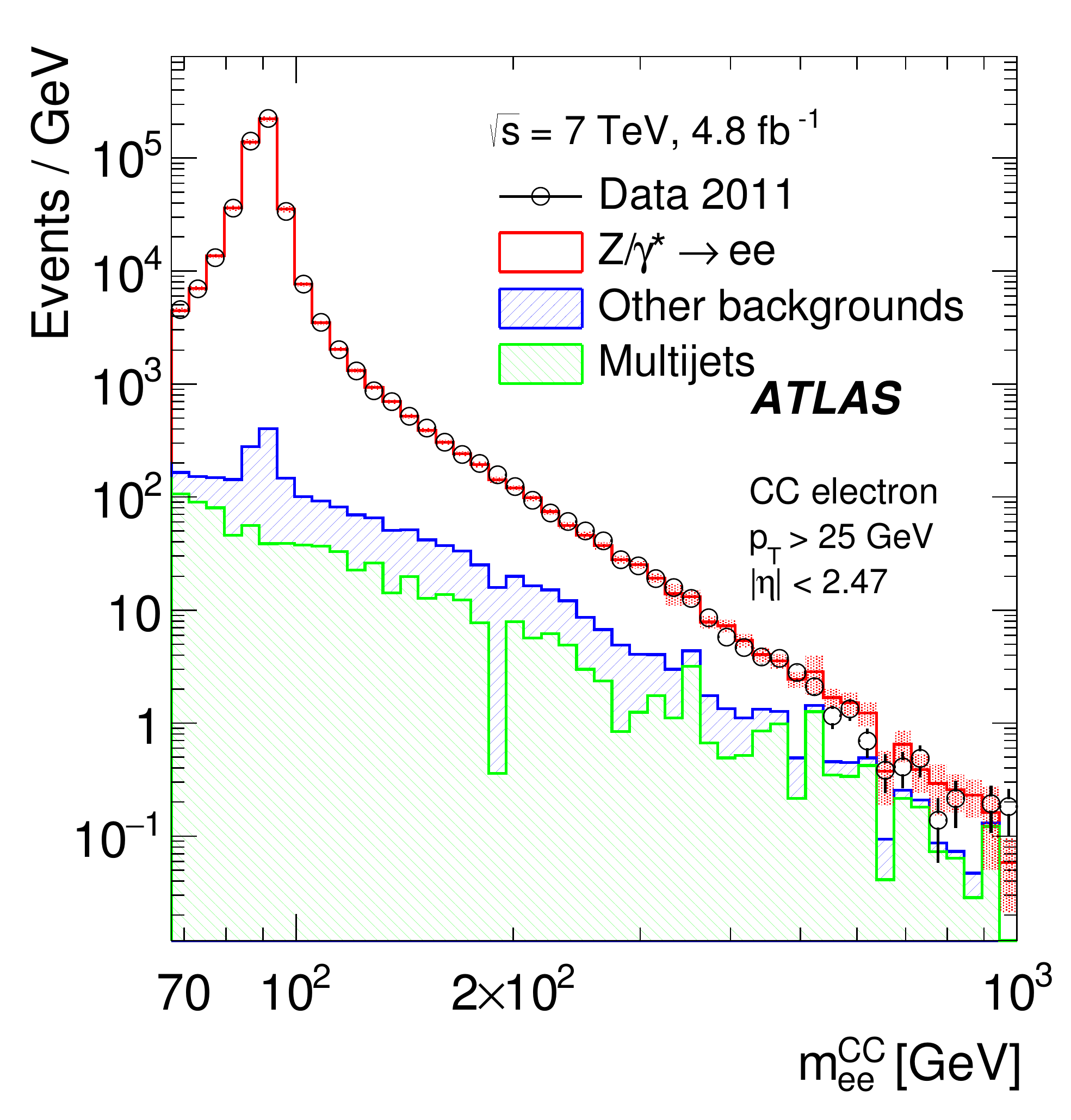}
      \label{fig:mass_CC}
      }
    \subfigure[]{      \includegraphics[width=0.444\textwidth,angle=0]{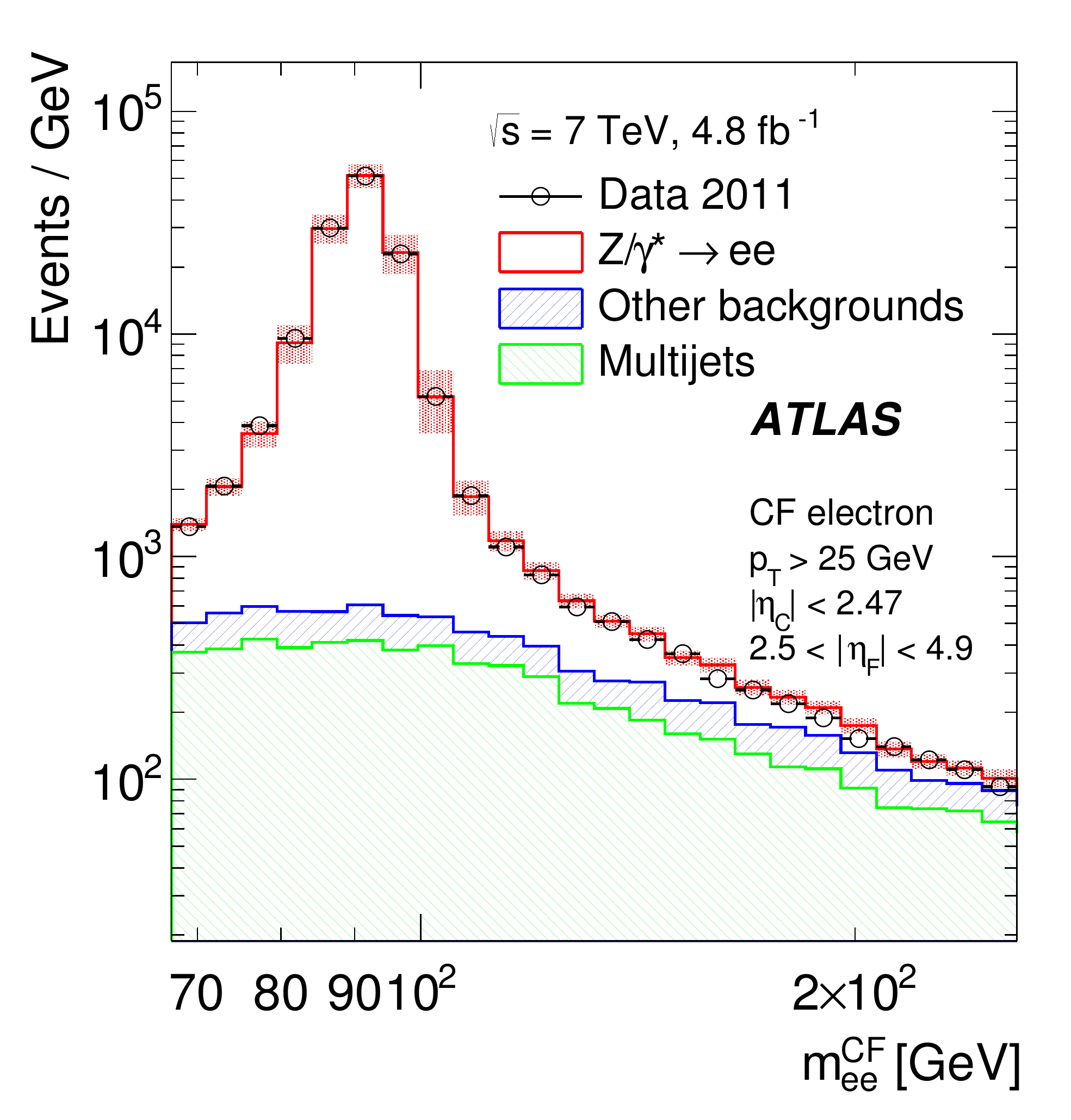}
      \label{fig:mass_CF}
      }
    \subfigure[]{      \includegraphics[width=0.444\textwidth,angle=0]{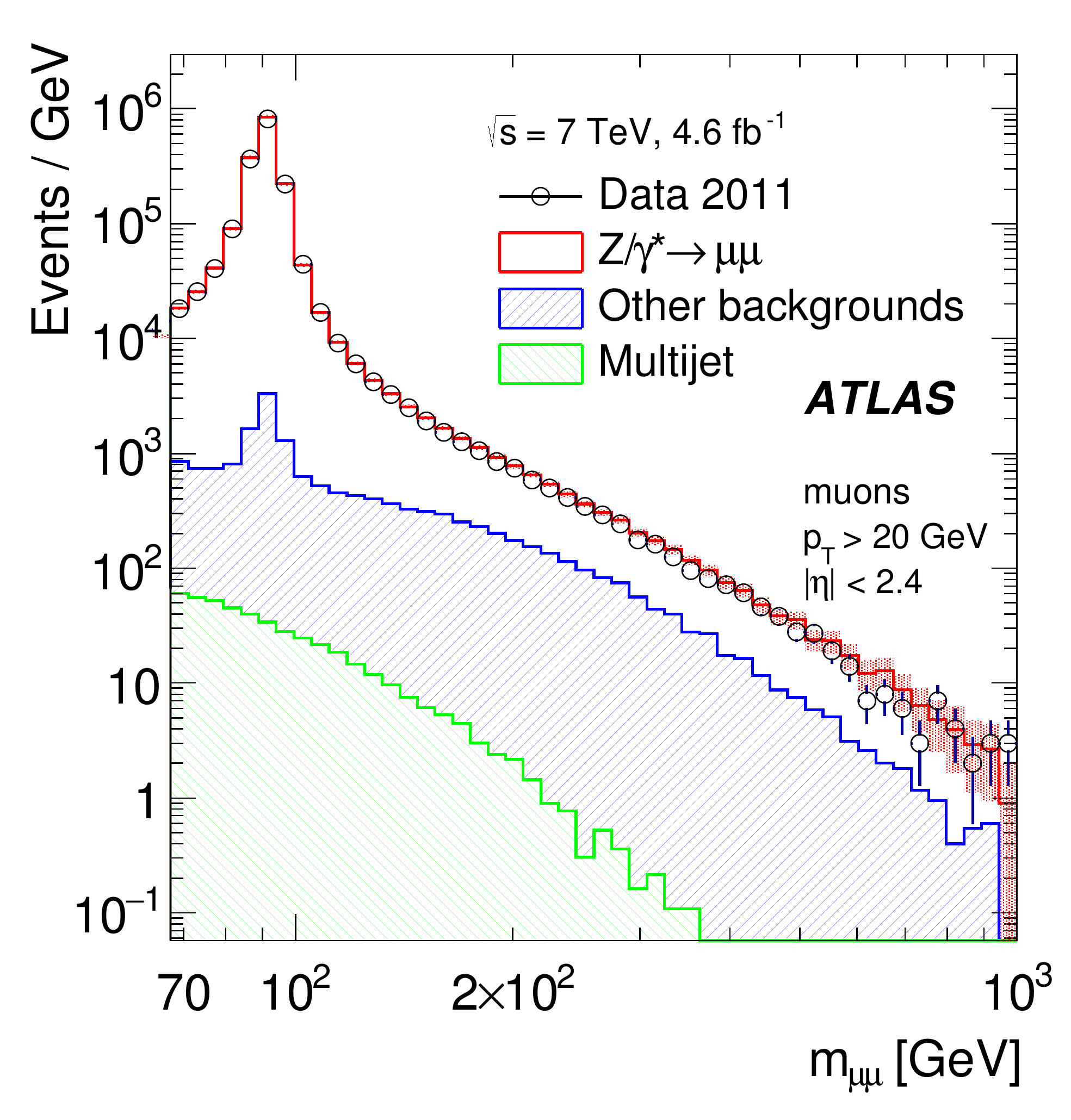}
      \label{fig:mass_mu}
      }
    \caption{
    Dilepton invariant mass distributions obtained from the event selections described in the text, for the
(a) CC electron, (b) CF electron and (c) muon channels. Data are shown by open circles and the total expectation is shown as a line
with a band representing the total uncertainty (statistical and systematic added in quadrature). The data-driven estimate for the
multi-jet background and the simulation-based estimates for all other backgrounds are shown by the shaded areas.  }
       \label{fig:mass}
  \end{center}
\end{figure}

\begin{figure}[p!]
  \begin{center}
    \subfigure[]{      \includegraphics[width=0.43\textwidth,angle=0]{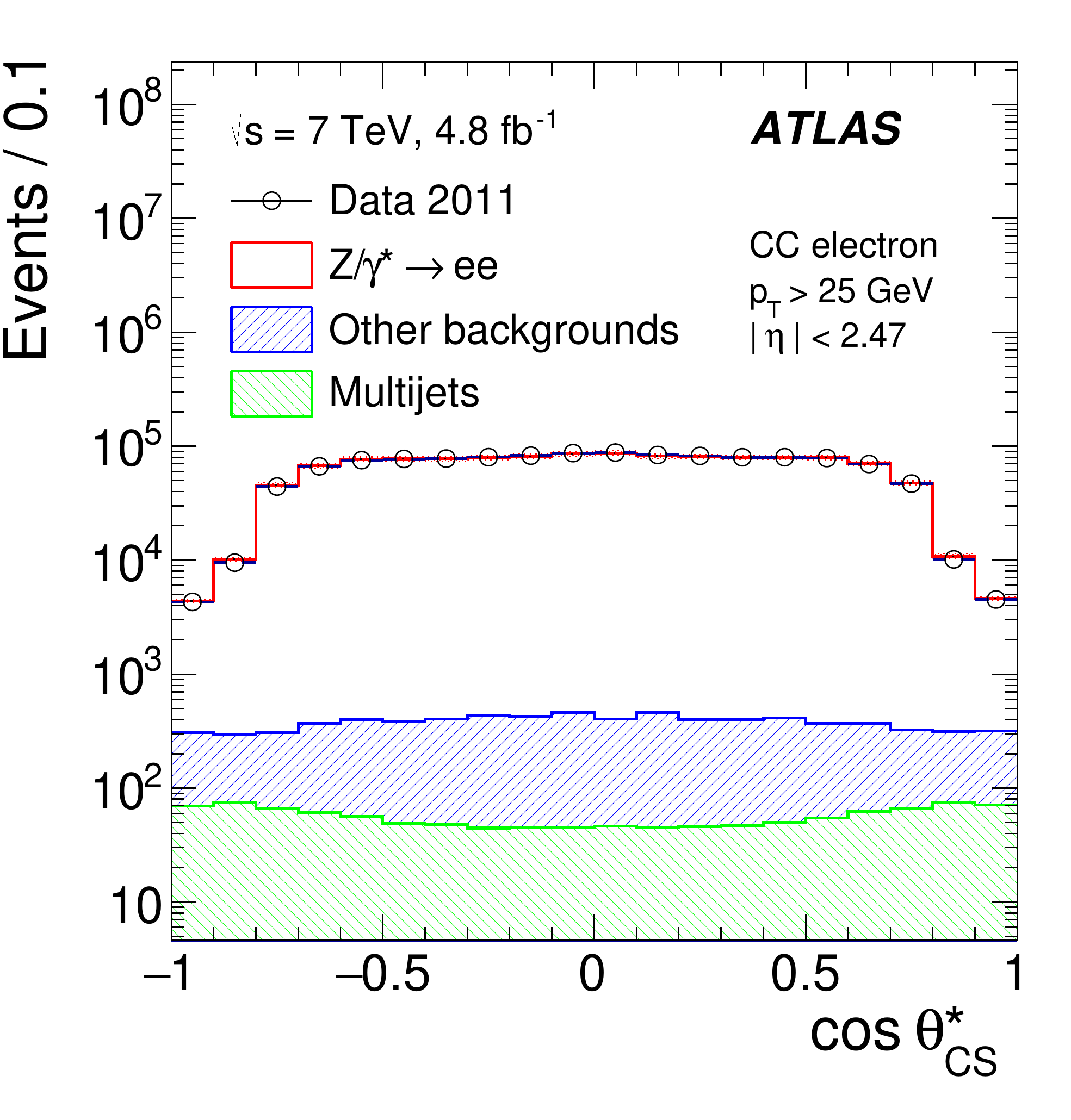}
      \label{fig:cos_CC}
      }
    \subfigure[]{      \includegraphics[width=0.43\textwidth,angle=0]{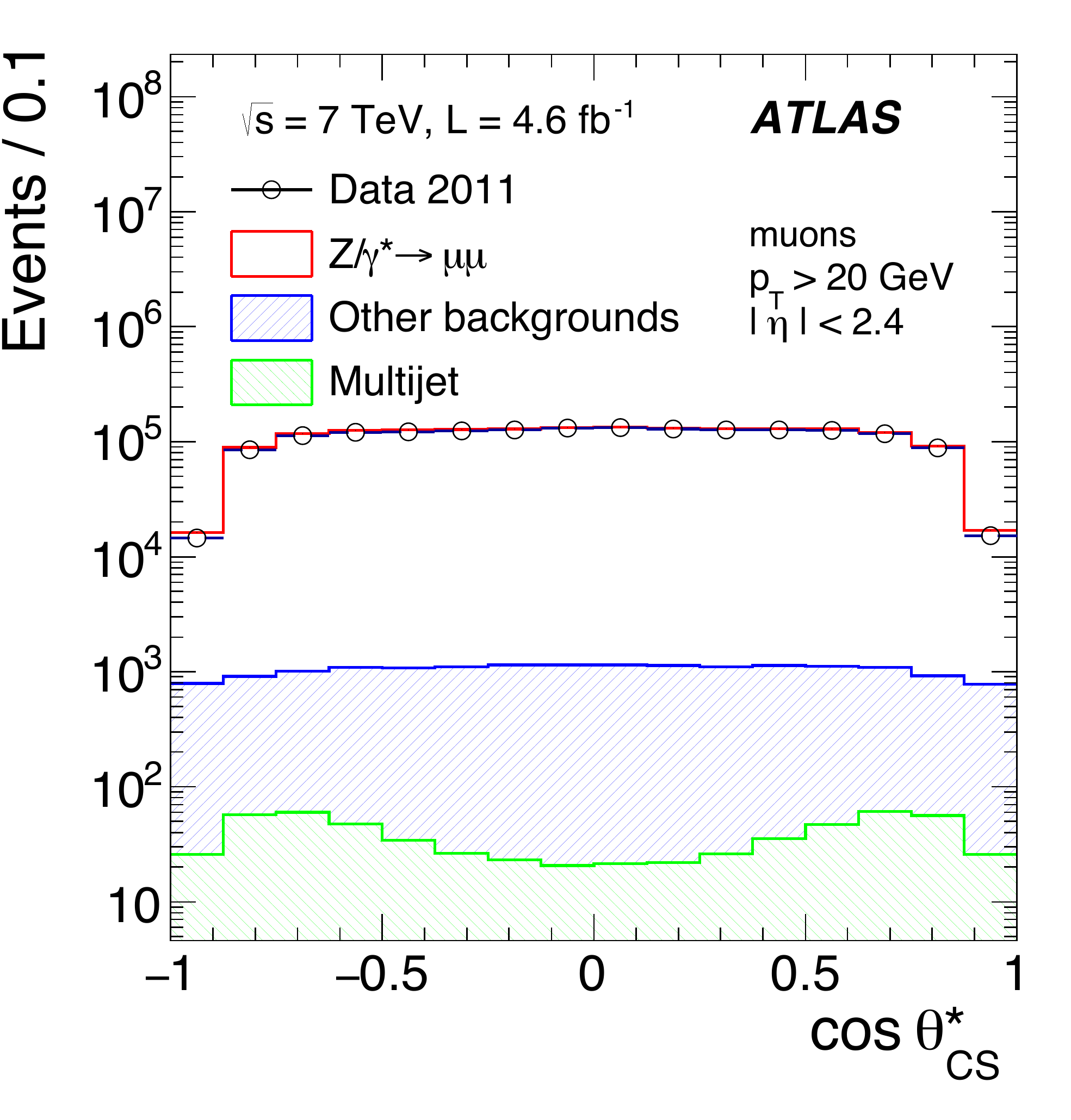}
      \label{fig:cos_mu}
      }
    \subfigure[]{      \includegraphics[width=0.43\textwidth,angle=0]{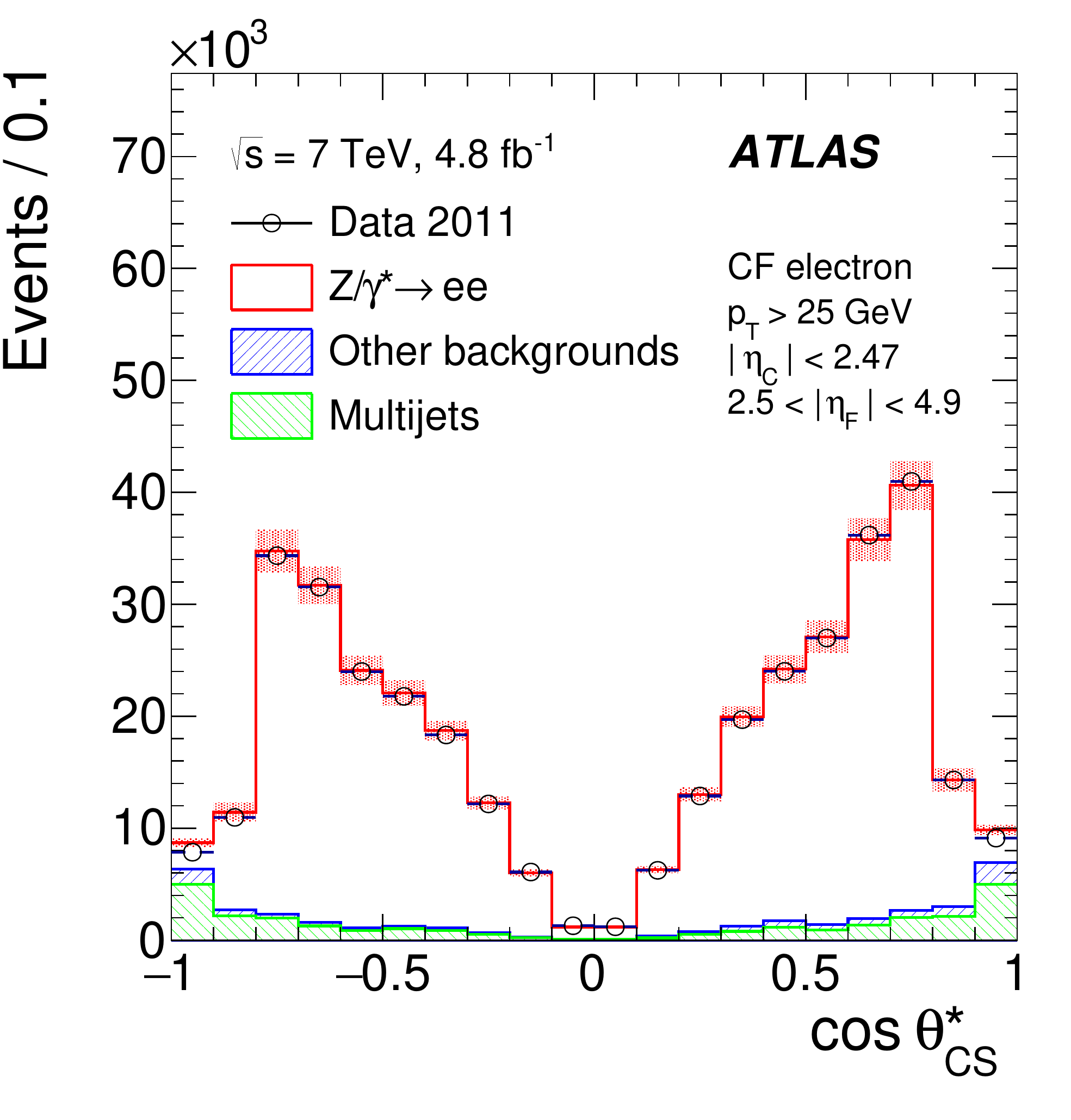}
      \label{fig:cos_CF}
      }
      \subfigure[]{      \includegraphics[width=0.43\textwidth,angle=0]{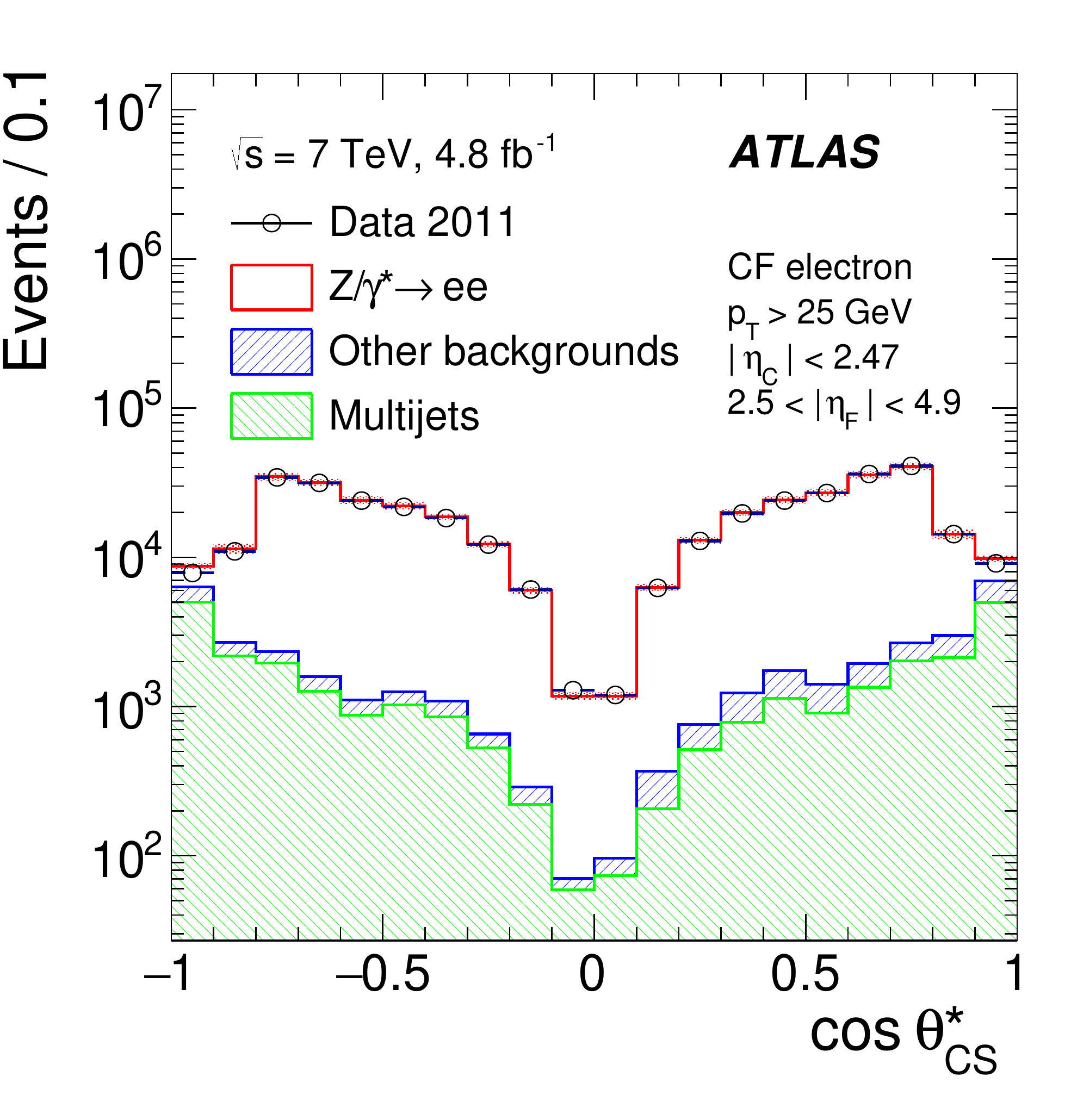}
      \label{fig:cos_CF_log}
      }
    \caption{Distributions of the cosine of the polar angle in the Collins--Soper frame (\cosThCS) obtained from the event selections described in the text, for the
(a) CC electron and (b) muon channels. The corresponding distribution for the CF electron channel is shown using both (c) a linear and (d) a logarithmic scale.
Data are shown by open circles and the total expectation is shown as a line
with a band representing the total uncertainty (statistical and systematic added in quadrature). The data-driven estimate for the
multi-jet background and the simulation-based estimates for all other backgrounds are shown by the shaded areas. }
       \label{fig:cos}
  \end{center}
\end{figure}

\Fignames~\ref{fig:mass} and \ref{fig:cos} show the $m_{\ell \ell}$ and \cosThCS~distributions
of events in the three channels. The total numbers of selected events are
1.2$\times$10$^6$, 0.35$\times$10$^6$ and 1.7$\times$10$^6$ for the CC electron, CF electron and muon channels respectively. In the region close to the
$Z$ peak, the background contamination is estimated to be less than 1\% for the muon and CC electron channels,
and about 5\% for the CF electron channel.
The background contributions in the muon and CC electron channels increase to about 5\% and 16\%
in the low- and high-mass regions, respectively. The CF electron channel has a background contamination of about
30\% in the low-mass region.
Agreement between data and simulation is observed within the uncertainties over the whole
invariant mass range and also in the \cosThCS~distributions.
These uncertainties contain both the statistical and systematic components and include the
effects of multiple $pp$ collisions occurring in the same or in neighbouring bunch crossing (pileup), energy/momentum scale and resolution, trigger efficiency, misalignment of the inner detectors, data-driven background estimates, and PDFs.
Details are given in \sectionname~\ref{sec:afbmeas}.

\Figname~\ref{fig:cos_CF} highlights the \cosThCS~distribution for the CF electron channel
to better illustrate the reduced impact of dilution: the forward-backward asymmetry is large enough to be observed directly from the plot.
Some differences between data and simulation are observed in the lowest and highest bins in \cosThCS. As a cross check, the analysis was repeated
excluding the bins in question and the impact on the \Afb~and \wma results was found to be negligible.

\section{Measurement of \texorpdfstring{\Afb}{AFB}~}
\label{sec:afbmeas}

For each invariant mass bin, the \Afb\ value is obtained from the corresponding \cosThCS\ distribution by
measuring the numbers of forward and backward events:

\[ \Afb = \frac{N_{\cosThCS\geq 0}-N_{\cosThCS<0}}{N_{\cosThCS\geq 0}+N_{\cosThCS<0}}. \]

\begin{figure}[p!]
   \begin{center}
    \subfigure[]{               \includegraphics[width=0.45\textwidth,angle=0]{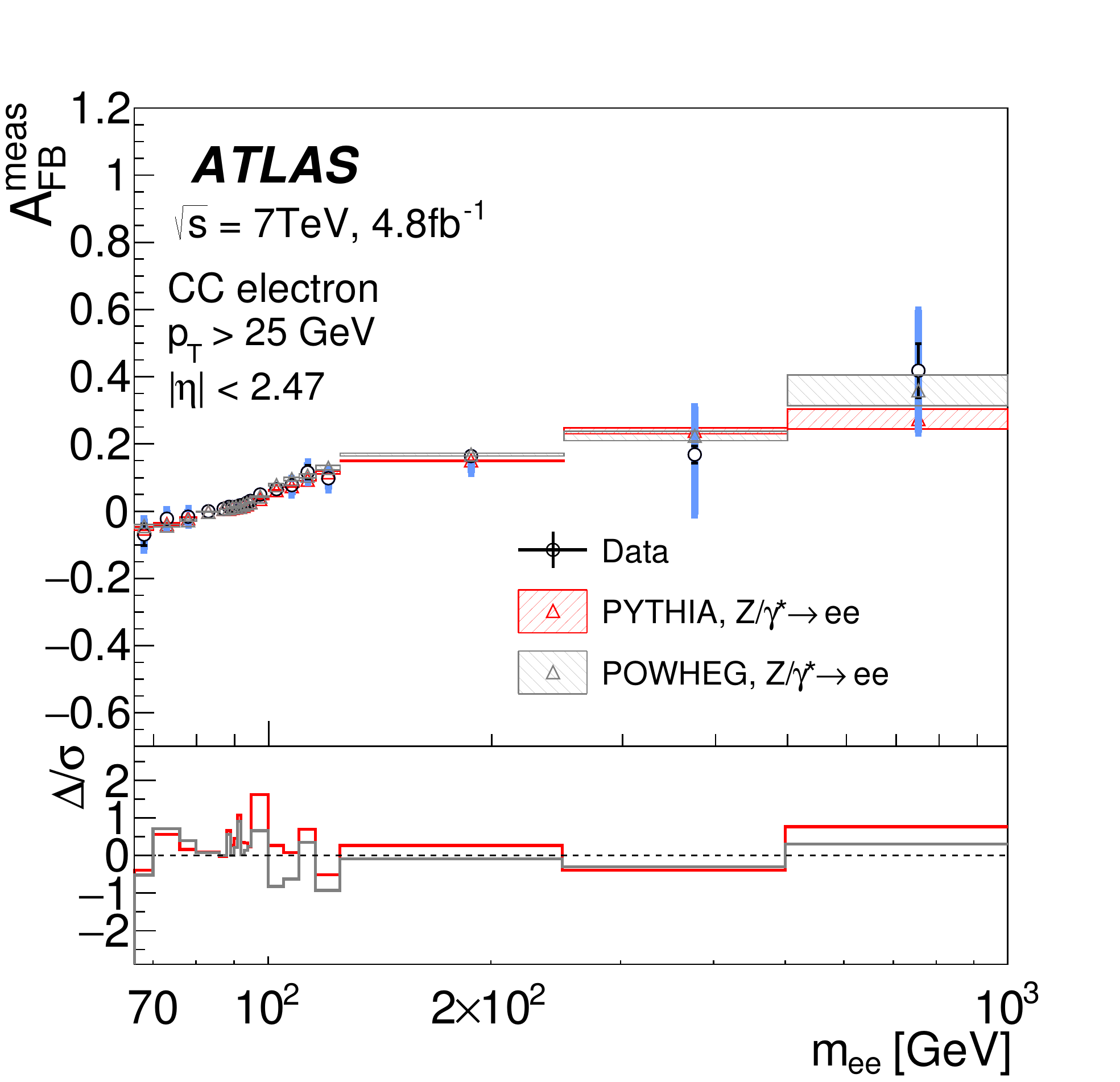}
               \label{fig:ele:rawAFB_CC}
      }
   \subfigure[]{               \includegraphics[width=0.45\textwidth,angle=0]{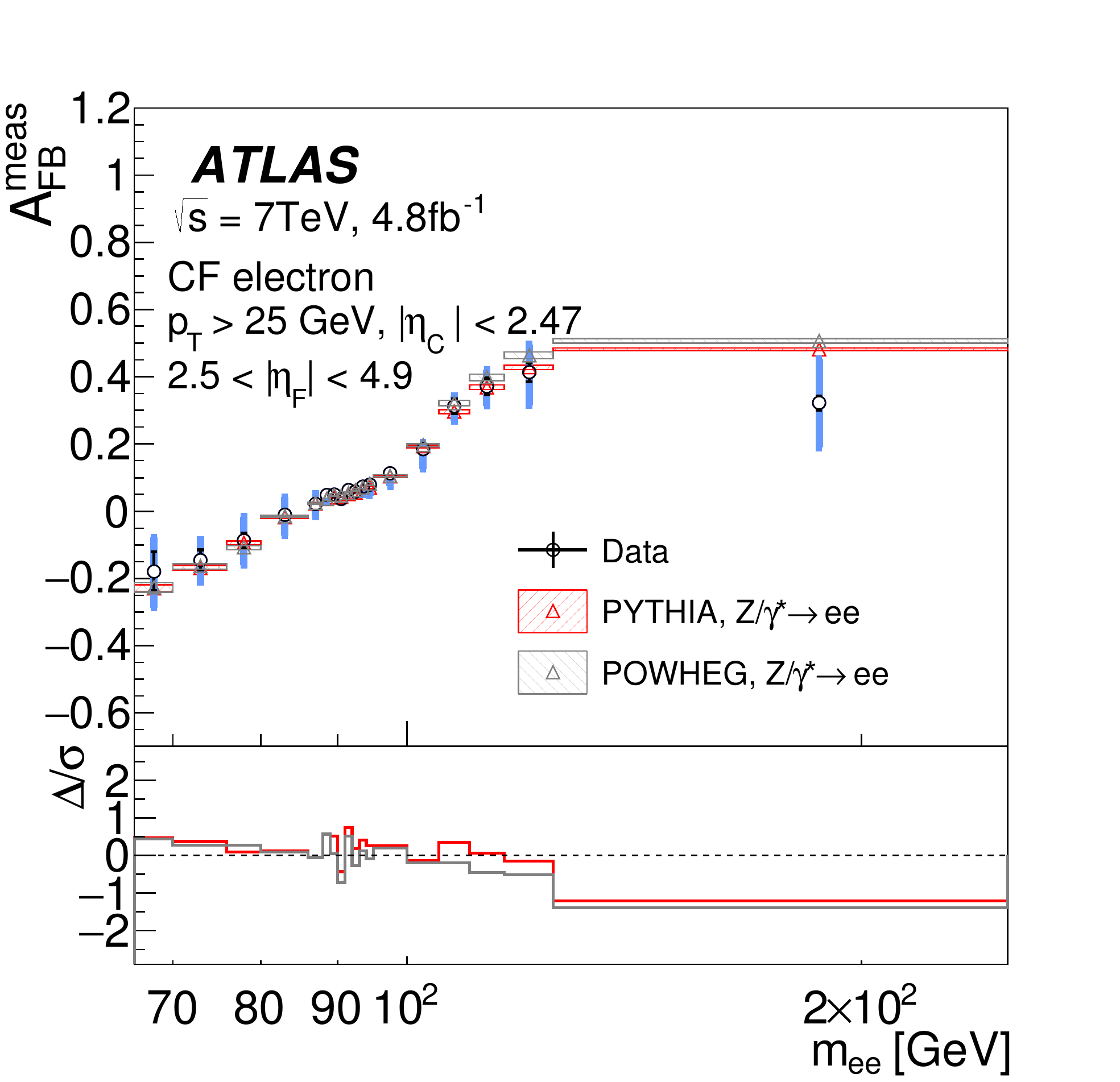}
               \label{fig:ele:rawAFB_CF}
      }
   \subfigure[]{              \includegraphics[width=0.45\textwidth,angle=0]{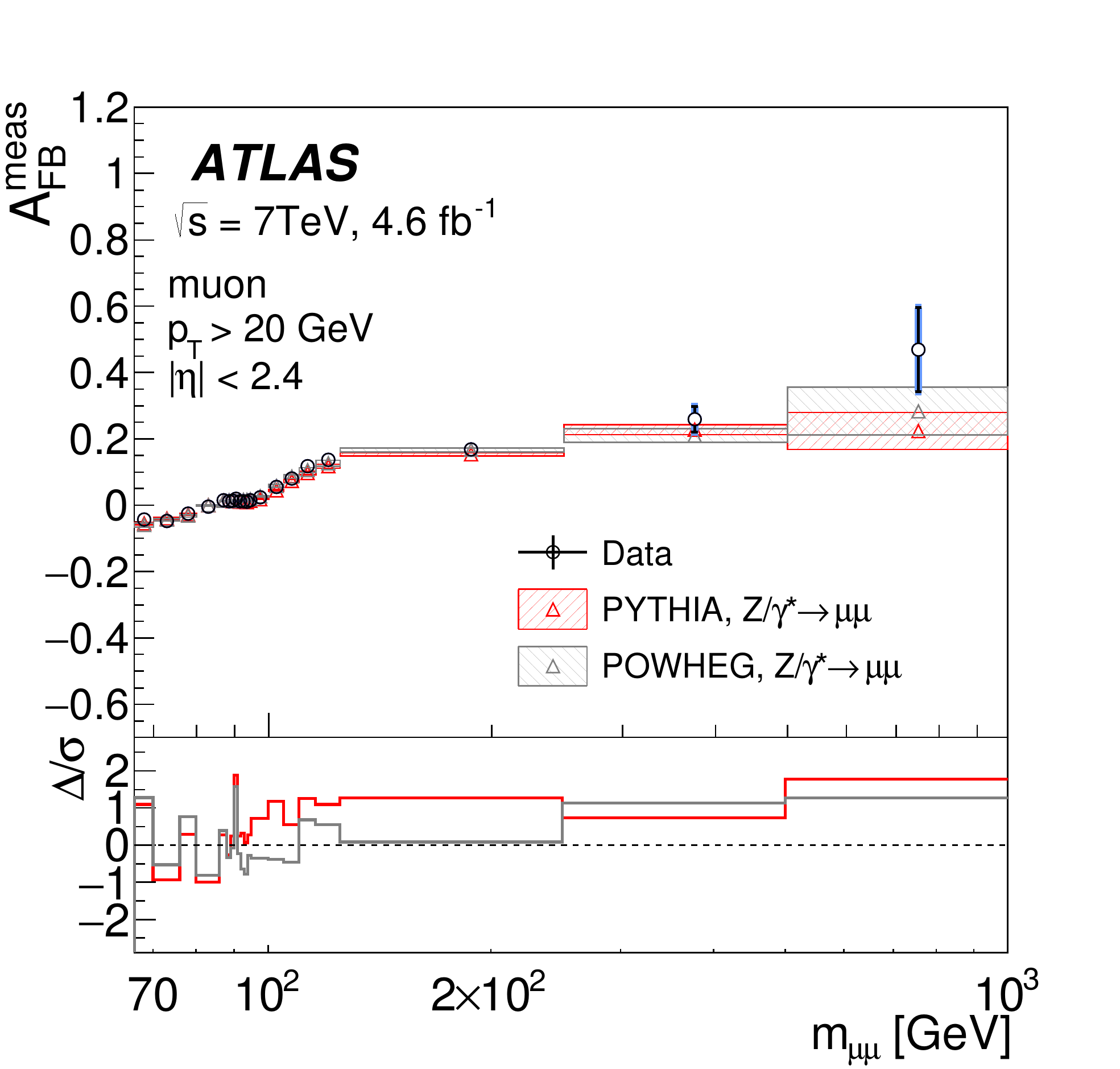}
                \label{fig:mu:rawAFB}
      }
   \caption{Detector-level forward-backward asymmetry ($\Afb^{\textrm{meas}}$) values as a function of the dilepton
   invariant mass for the (a) CC electron, (b) CF electron and
   (c) muon channels, after background subtraction. For the data, the black inner error bars represent
   the statistical component and the lighter outer error bars the total error (statistical and systematic
   added in quadrature). The boxed shaded regions for the MC expectations represent only the statistical
   uncertainty;
   theoretical uncertainties for MC are included
   in the systematic uncertainties on the data.
   The lower panel of each plot shows the pull value ($\Delta/\sigma$) for each mass bin, where $\Delta$ is the difference between
   data and simulation and $\sigma$ is the sum in quadrature of the data and simulation uncertainties.}
   \label{fig:all:raw}
     \end{center}
 \end{figure}

\begin{figure}[p!]
   \begin{center}
    \subfigure[]{               \includegraphics[width=0.45\textwidth,angle=0]{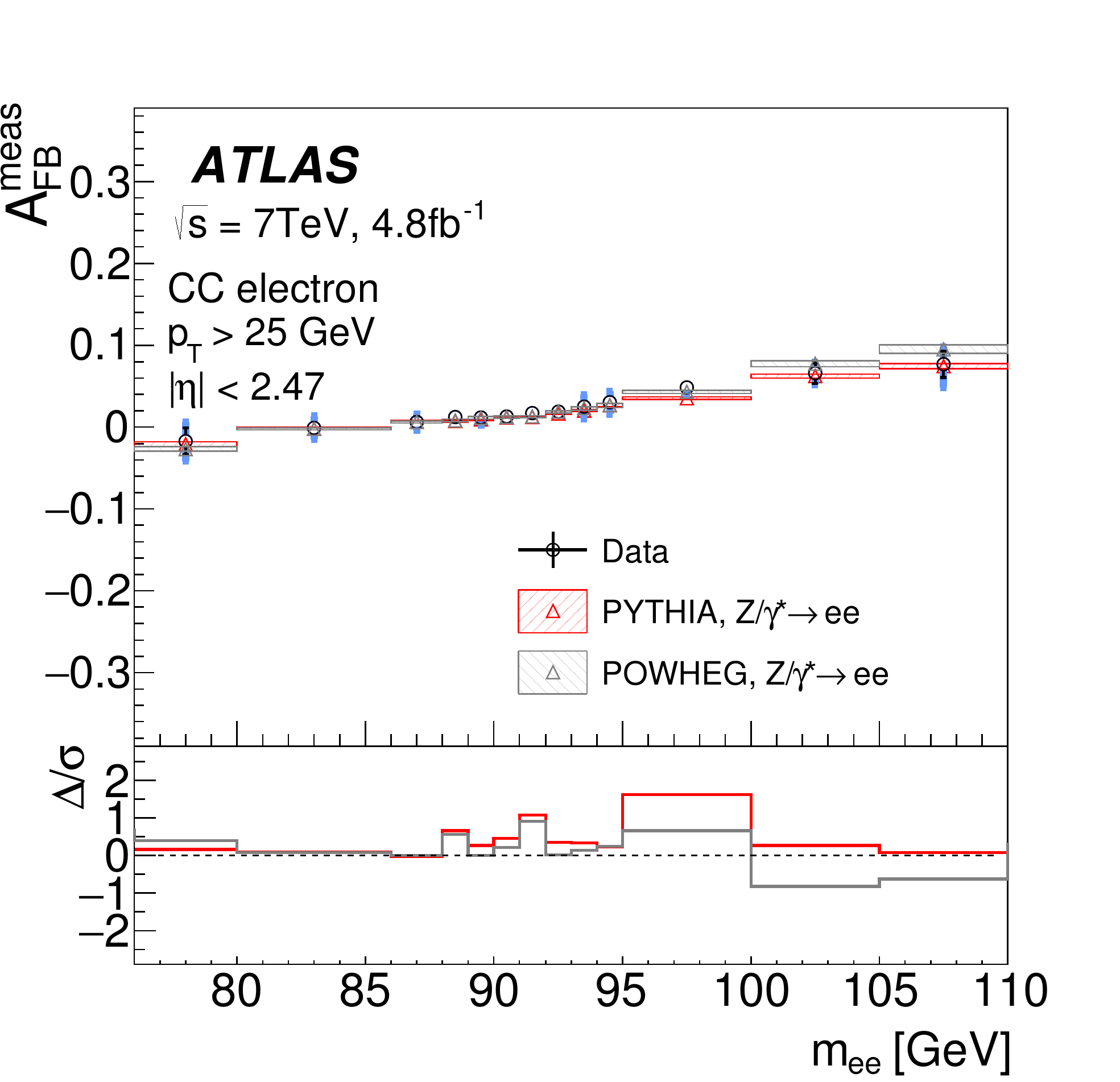}
               \label{fig:ele:rawAFBzoom_CC}
      }
   \subfigure[]{               \includegraphics[width=0.45\textwidth,angle=0]{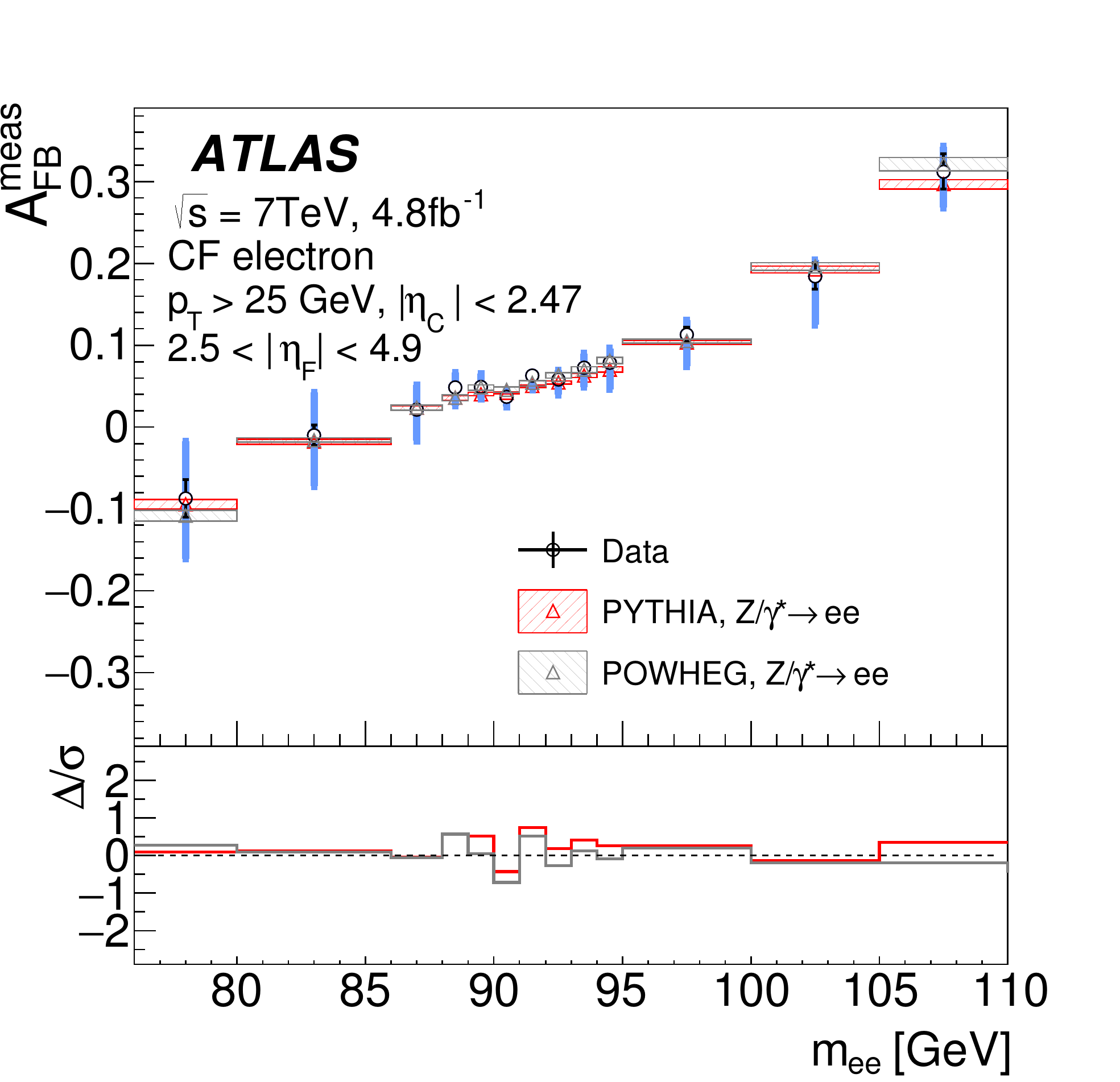}
               \label{fig:ele:rawAFBzoom_CF}
      }
   \subfigure[]{               \includegraphics[width=0.45\textwidth,angle=0]{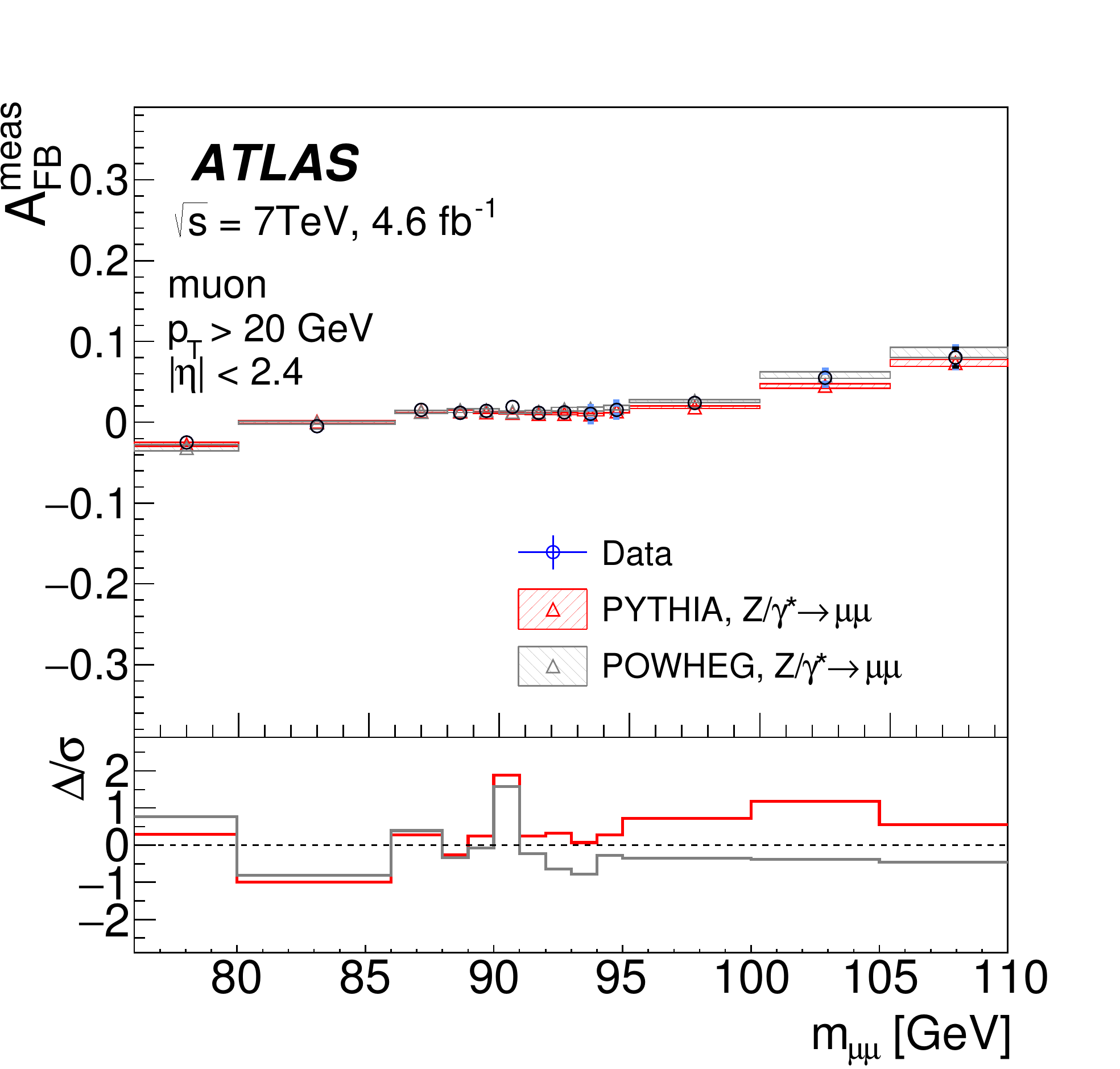}
                \label{fig:mu:rawAFBzoom}
      }
   \caption{Detector-level forward-backward asymmetry ($\Afb^{\textrm{meas}}$) values as a function of the dilepton
   invariant mass for the (a) CC electron, (b) CF electron and (c) muon channels in
	a narrow region around the $Z$ pole, after background subtraction.
	For the data, the black inner error bars represent
   the statistical component and the lighter outer error bars the total error (statistical and systematic
   added in quadrature). The boxed shaded regions for the MC expectations represent only
   the statistical uncertainty; theoretical uncertainties for MC are included
   in the systematic uncertainties on the data.
   The lower panel of each plot shows the pull values ($\Delta/\sigma$, as defined for \figname~\ref{fig:all:raw}).}

   \label{fig:all:rawzoom}
     \end{center}
 \end{figure}

\noindent For comparison, expected \Afb\ values are calculated from both the PYTHIA
and POWHEG samples described in \sectionname~\ref{sec:dataMC}. Background contributions
are subtracted from the number of forward and backward events measured at detector-level.
Some background contributions, such as multi-jet events, display no asymmetry and hence dilute the
measured asymmetry.
Other background contributions, such as
$\ttbar$, display an asymmetry.
The detector-level asymmetry values after background subtraction ($\Afb^{\textrm{meas}}$) in the electron and muon
channels are shown in \figname~\ref{fig:all:raw} as a function of the invariant mass of the
lepton pair\footnote{Numerical values of all results are available in HepData \cite{Hepdata}.}.
Good agreement between data and simulation is observed.
\Figname~\ref{fig:all:rawzoom}~shows the same information in a narrower mass range around
the $Z$ pole.
\label{sec:afbdef}

The asymmetry values $\Afb^{\textrm{meas}}$ are unfolded from detector level to particle
level ($\Afb^{\textrm{obs}}$), to allow comparisons with theoretical predictions.
The unfolding procedure corrects for effects collectively referred to as
'mass-bin migration' (MBM) as described below.
\begin{itemize}
\item
\textit{Detector effects}: the finite resolution of the detector, as well as lepton reconstruction efficiencies, deform
the measured \zgamma\ line shape and the dependence of the asymmetry values on the dilepton mass
with respect to what one would measure with an ideal apparatus covering the same kinematic range.
\item \textit{QED radiative corrections}: radiative corrections~\cite{QED}, or real photon emission in the final-state
(FSR), deform
the shape of the dilepton invariant mass distribution. This deformation is particularly pronounced below the $Z$ peak.
The events are moved from the $Z$ peak (i.e.\ expected $\Afb$ positive and small)
towards smaller values of
invariant mass, significantly reducing the magnitude of the observed $\Afb$ in the region
66\,GeV$<m_{\ell\ell}<m_Z$. In the high-mass region ($m_{\ell\ell}>m_Z$), the deformation due to
radiative corrections is still present, but is reduced in magnitude.
To account for these corrections, dileptons are unfolded to the pre-FSR state, referred to as `Born
level'.\\
\end{itemize}

The unfolding procedure is carried out using an iterative Bayesian
unfolding method~\cite{dagostini}, as implemented in the RooUnfold toolkit~\cite{roounfold}.
The response matrices are built from the PYTHIA signal sample and the number of iterations (ten) is chosen in such a way
as to optimize the result of closure tests on simulated samples.
Additional checks are performed to ensure that the use of the
PYTHIA LO generator for the unfolding does not bias the result.
Since FSR is a significant correction, an alternative real-photon emission generator is investigated,
using a simulated sample generated with SHERPA~\cite{sherpa}, which uses
a module called PHOTONS++~\cite{photons} for higher-order QED
corrections.
The impact of NLO electroweak (EWK) corrections on the response matrix are estimated by
reweighting the PYTHIA simulation to the prediction from the
HORACE MC event generator and redoing the unfolding.
In order to estimate NLO QCD effects on the $\Afb^{\textrm{obs}}$~values, a test is performed
using a simulated POWHEG sample as pseudo-data
and unfolding the asymmetry values using the PYTHIA-derived response matrices.
These studies all show that any biases are much smaller than the present statistical uncertainties of the measurement.

The systematic uncertainties on $\Afb^{\textrm{obs}}$ have contributions from the
sources discussed in the following.

\begin{figure}[p!]
  \begin{center}
    \subfigure[]{      \includegraphics[width=0.45\textwidth,angle=0]{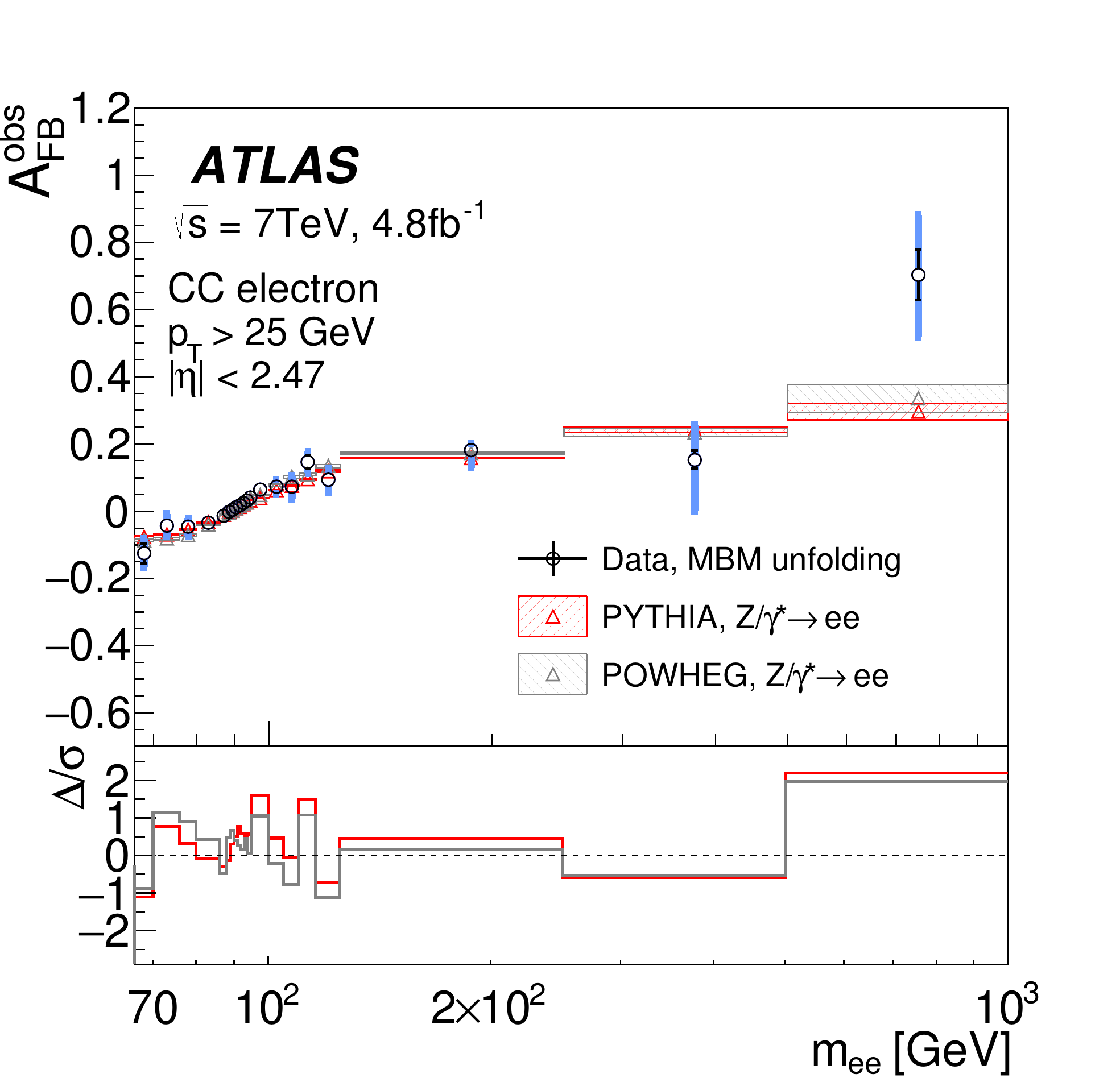}
      \label{fig:elemassunfold_CC}
      }
    \subfigure[]{      \includegraphics[width=0.45\textwidth,angle=0]{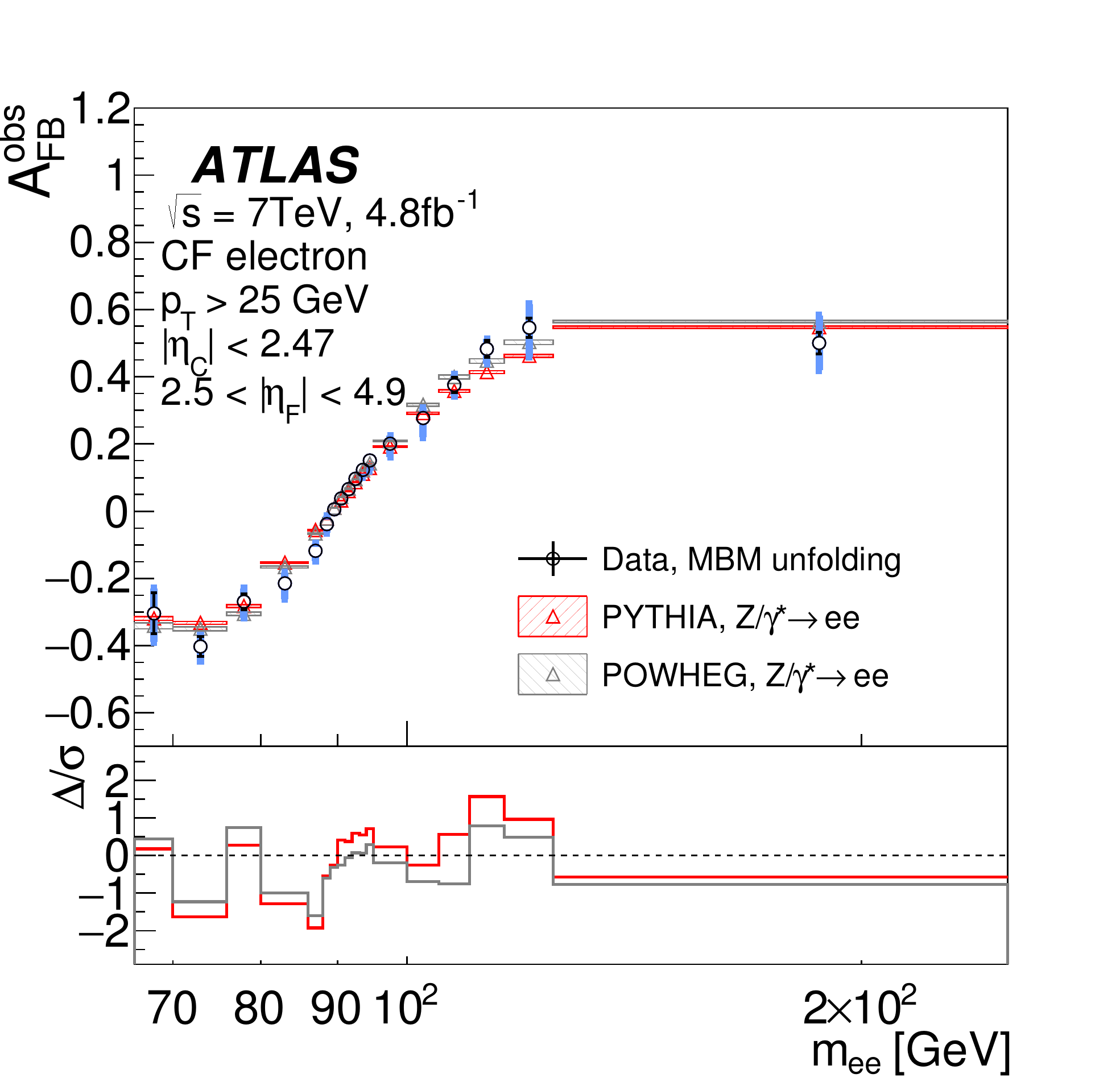}
      \label{fig:elemassunfold_CF}
      }
    \subfigure[]{      \includegraphics[width=0.45\textwidth,angle=0]{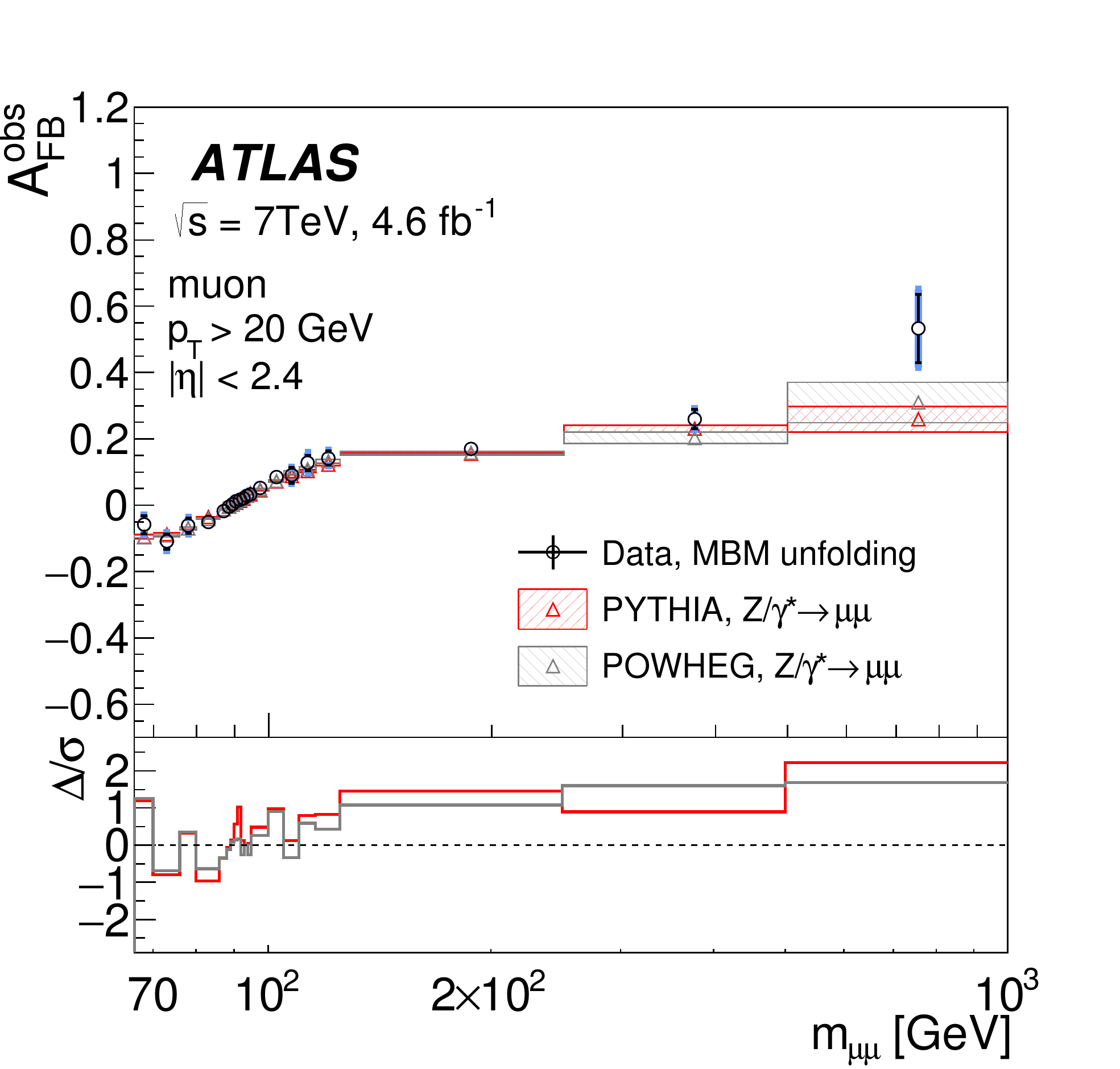}
      \label{fig:mumassunfold}
      }
    \caption{Forward-backward asymmetry ($\Afb^{\textrm{obs}}$) values as a function of the dilepton
   invariant mass for the (a) CC electron, (b) CF electron and (c) muon channels.
   Leptons are unfolded to Born level to account for mass bin migration, and the results
   are compared to truth-level MC information.
     For the data, the black inner error bars represent
   the statistical component and the lighter outer error bars the total error (statistical and systematic
   added in quadrature).
      The boxed shaded regions for the MC expectations represent only the statistical uncertainty; theoretical uncertainties are included in
   the systematic uncertainties on the data.
   The lower panel of each plot shows the pull values ($\Delta/\sigma$, as defined for \figname~\ref{fig:all:raw}).}
       \label{fig:all:massunfold}
  \end{center}
\end{figure}

\begin{itemize}
 \item \textit{Unfolding uncertainty}: estimated using a partially data-driven method.
 A set of weights is derived as a function of $m_{\ell \ell}$ and $\cosThCS$
 to reweight the $\Afb^{\textrm{meas}}$ values from simulation to the
 one observed in data. These weights
are applied to the generator-level asymmetry values. The response matrix used in the unfolding is
applied to the resulting values to fold and subsequently unfold them. Particular care is taken to make
the matrices used for the folding and the unfolding
statistically independent. The generator-level \Afb\ dependence on $m_{\ell \ell}$ is compared before and
after the fold--unfold operation, and the difference
is taken as an estimate of the uncertainty introduced by the unfolding.

\item \textit{Uncertainty due to finite size of the simulated event samples}: the statistical uncertainty on the response matrices
is propagated through the unfolding procedure.

\item \textit{Multi-jet background modelling}: in the CC electron channel, the difference between the two background
 estimation methods described in~\sectionname~\ref{sec:eventselection} is taken as the systematic
 uncertainty and
 is found to be negligible with respect to other uncertainties.
 In the CF electron channel,
 this uncertainty is estimated by comparing templates based on different electron isolation requirements.
 For the muon channel, the impact of this background (and its uncertainty) is negligible.

\item \textit{Other experimental systematic uncertainties}: these include the impact of pileup and of detector
alignment, as well as energy/momentum scale and resolution, and trigger and reconstruction efficiencies.
The associated systematic uncertainties are estimated following the prescriptions in Refs.~\cite{EurPhysJC72, muonPerformance, Alignment}.
The uncertainties related to energy scaling and resolution are among the largest contributions to the total error in all three channels,
since they result in both a shifting and a broadening of the invariant mass peak, causing events to migrate between mass bins.

\item \textit{PDF uncertainties}: the CT10 PDF set~\cite{CTEQ}, which provides a reliable uncertainty
estimate and is widely used in ATLAS,
is also used here to estimate the PDF uncertainty.
 Its eigenvectors are used and the result quoted at 68\%~confidence level. For each error set, the MC signal sample is reweighted, the response
 matrices are recalculated and the unfolding is repeated.
 This contribution is found to be small when unfolding only mass-bin migration effects.

 \end{itemize}
The magnitudes of the systematic uncertainties on the $\Afb^{\textrm{obs}}$ values are summarized in \tabname~\ref{tab:afbunc},
for three invariant mass regions.
\Figname~\ref{fig:all:massunfold} shows the $\Afb^{\textrm{obs}}$ values obtained from leptons unfolded to
Born level for all three channels.
Expectations from PYTHIA and POWHEG are in good agreement with the measured values, as illustrated
in the pull distribution at the bottom of each plot.

 \begin{table}[hbtp]
  \centering
  \begin{tabular}{l|c|c|c}
    \hline\hline
    \multicolumn{4}{c}{\textbf{CC electrons}}  \\
    Uncertainty          	& 66--70\,GeV     & 70--250\,GeV        & 250--1000\,GeV    \\ \hline\hline

    Unfolding 				&  $\sim$1$\times$10$^{-2}$	&  (2--5)$\times$10$^{-3}$	&  	$\sim$4$\times$10$^{-4}$	   			\\ \hline
    Energy scale/resolution	&  $\sim$7$\times$10$^{-3}$	&  (0.5--2)$\times$10$^{-3}$	& 	$\sim$2$\times$10$^{-2}$	          		\\ \hline
    MC statistics        				&  $\sim$5$\times$10$^{-3}$	&  (0.1--1)$\times$10$^{-3}$	&   	(3--20)$\times$10$^{-3}$	 	\\\hline
    PDF					&  $\sim$2$\times$10$^{-3}$	&  (1--8)$\times$10$^{-4}$	& 	(0.7--3)$\times$10$^{-3}$		\\\hline
    Other					&  $\sim$1$\times$10$^{-3}$	&  (0.1--2)$\times$10$^{-3}$	& 	(5--9)$\times$10$^{-3}$	          \\ \hline\hline

    \multicolumn{4}{c}{\textbf{CF electrons}} \\
    Uncertainty          	& 66--70\,GeV     & 70--250\,GeV        & 250--1000\,GeV    \\ \hline\hline

    Unfolding 				&  $\sim$2$\times$10$^{-2}$   	&	(0.5--2)$\times$10$^{-2}$		&  	--	   	\\ \hline
    Energy scale/resolution	&  $\sim$1$\times$10$^{-2}$    & 	(0.5--7)$\times$10$^{-2}$	 	& 	--	          \\ \hline
    MC statistics         				&  $\sim$1$\times$10$^{-2}$ 	&   	(1--7)$\times$10$^{-3}$ 		&   	--	 	\\\hline
    Background				&  $\sim$3$\times$10$^{-2}$ 	&	(0.5--1)$\times$10$^{-2}$ 	&	--		\\ \hline
    PDF					&  $\sim$4$\times$10$^{-3}$ 	& 	(2--6)$\times$10$^{-4}$ 		& 	--		\\\hline
    Other					&  $\sim$1$\times$10$^{-3}$ 	& 	(1--5)$\times$10$^{-4}$ 		& 	--	          \\ \hline\hline

    \multicolumn{4}{c}{\textbf{Muons}}\\
    Uncertainty	& 	66--70\,GeV & 70--250\,GeV& 250--1000\,GeV \\ \hline\hline

    Unfolding 				&  $\sim$1$\times$10$^{-2}$	&  (1--4)$\times$10$^{-3}$		&  	$\sim$5$\times$10$^{-4}$	   			\\ \hline
    Energy scale/resolution	&  $\sim$8$\times$10$^{-3}$  	&  (3--6)$\times$10$^{-3}$         	& 	$\sim$5$\times$10$^{-3}$	          \\ \hline
    MC statistics         				&  $\sim$5$\times$10$^{-3}$	&  (0.1--1)$\times$10$^{-3}$		&   	(2--30)$\times$10$^{-3}$	 	\\\hline
    PDF					&  $\sim$2$\times$10$^{-3}$	&  (1--8)$\times$10$^{-4}$		& 	(0.3--3)$\times$10$^{-3}$		\\\hline
    Other					&  $\sim$1$\times$10$^{-3}$	&   (0.5--1)$\times$10$^{-3}$         	& 	(3--10)$\times$10$^{-3}$	          \\ \hline\hline

	\end{tabular}
	\caption{Absolute systematic uncertainties on the $\Afb^{\textrm{obs}}$ values, after unfolding for
	mass-bin migration. Approximate values
	in three invariant mass intervals are given.}
	\label{tab:afbunc}
\end{table}

\subsection{Correcting for dilution}
\begin{figure}[p!]
  \begin{center}
    \subfigure[]{      \includegraphics[width=0.45\textwidth,angle=0]{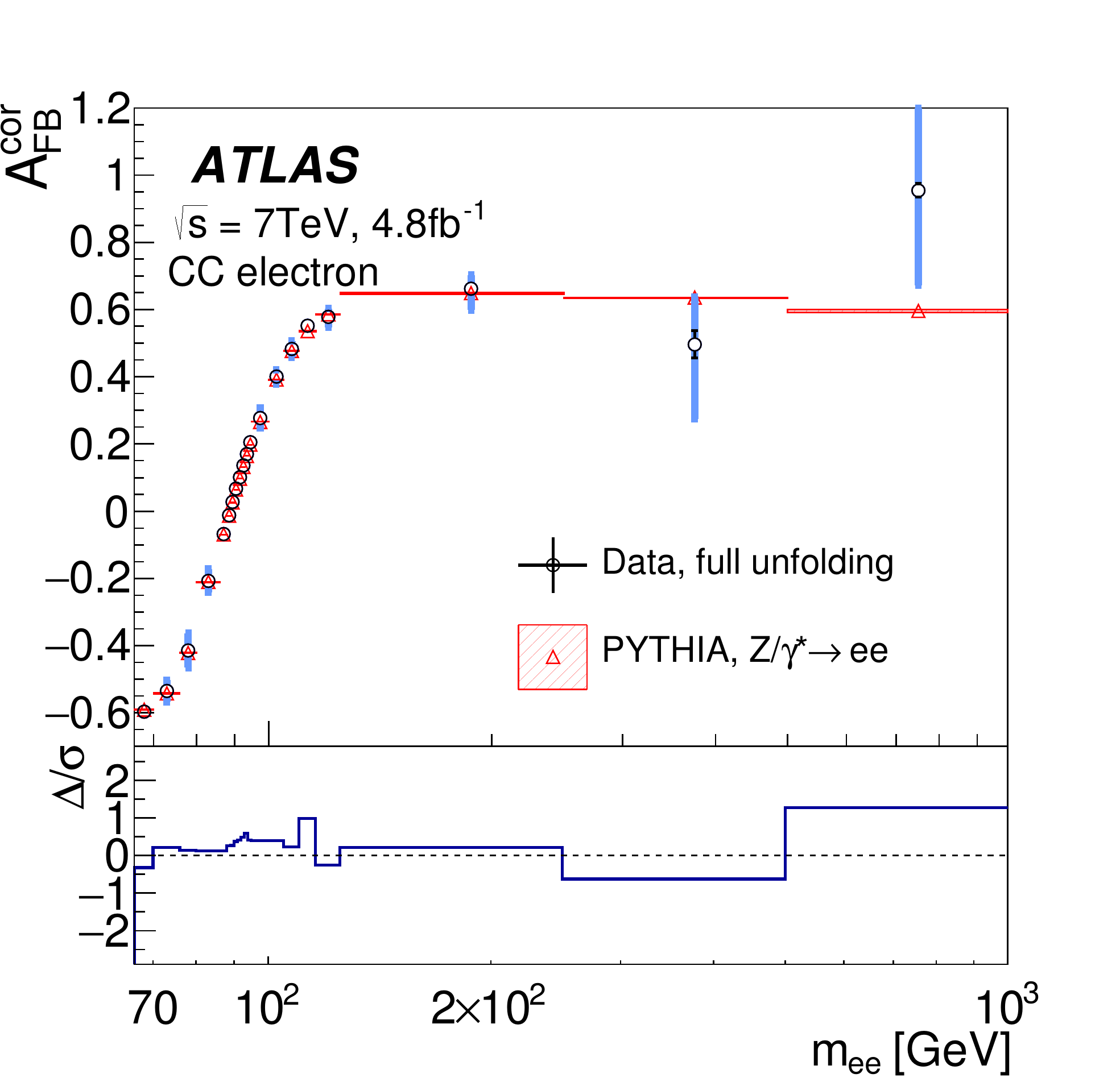}
      \label{fig:elefullunfold_CC}
      }
    \subfigure[]{      \includegraphics[width=0.45\textwidth,angle=0]{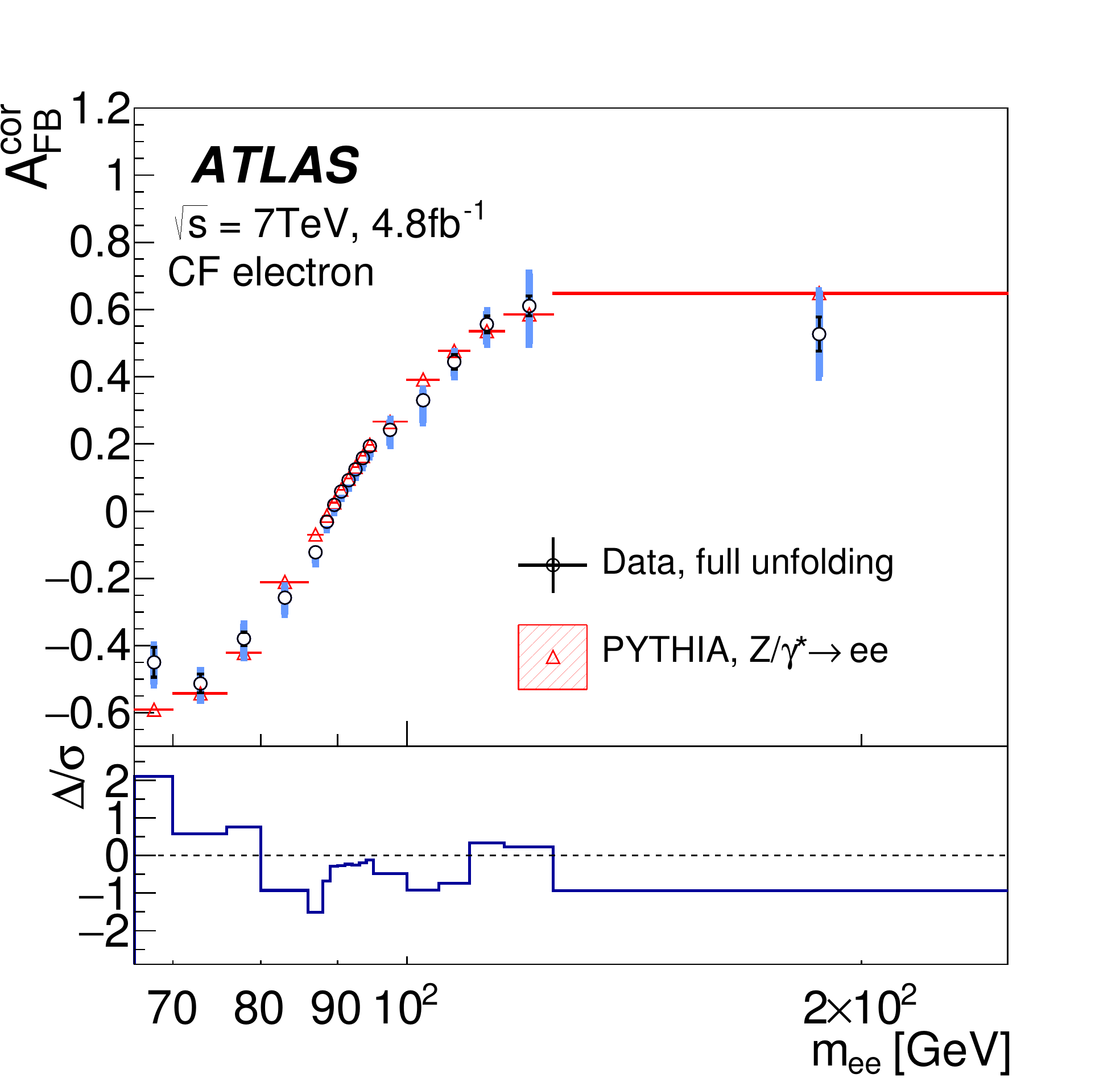}
      \label{fig:elefullunfold_CF}
      }
    \subfigure[]{      \includegraphics[width=0.45\textwidth,angle=0]{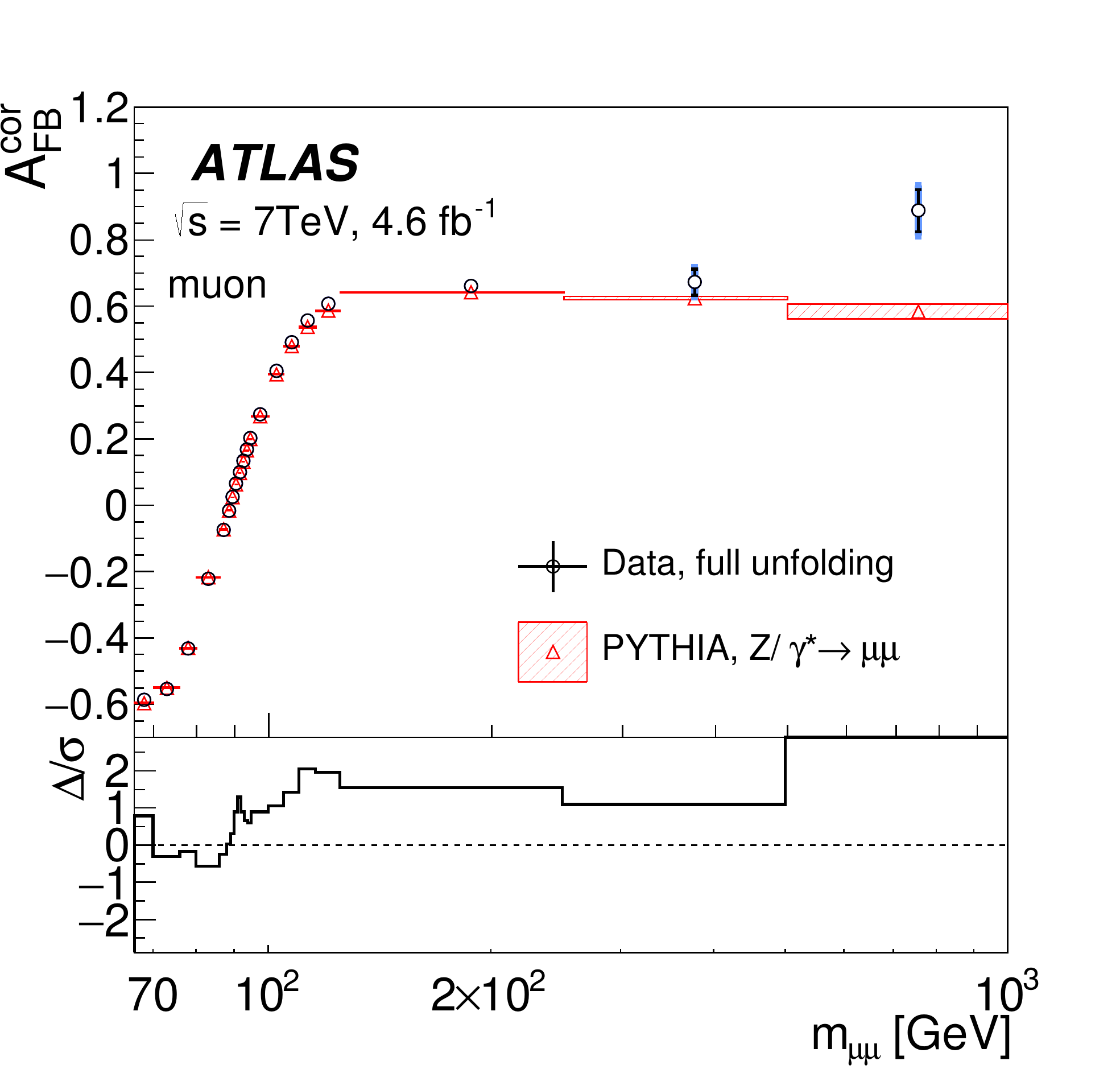}
      \label{fig:mufullunfold}
      }
    \caption{Forward-backward asymmetry ($\Afb^{\textrm{cor}}$) values as a function of the dilepton
   invariant mass for the (a) CC electron, (b) CF electron and (c) muon channels.
   Leptons are unfolded to Born level to account for mass bin migration and dilution effects are corrected.
   The measurement is extrapolated to the full phase space, and the results
   are compared to truth-level MC information.
   For the data, the black inner error bars represent
   the statistical component and the lighter outer error bars the total error (statistical and systematic
   added in quadrature). The boxed shaded regions for the MC expectations represent only the statistical uncertainty; theoretical uncertainties are included
   in the data systematic uncertainties.
   The lower panel of each plot shows the pull values ($\Delta/\sigma$, as defined for \figname~\ref{fig:all:raw}).}
       \label{fig:all:fullunfold}
  \end{center}
\end{figure}

A similar unfolding procedure is used to further correct the $\Afb^{\textrm{obs}}$
values
to remove dilution effects, which occur when the wrong
choice is made for the direction of the quark.
The unfolding
for dilution and the extrapolation from the detector acceptance to the full phase space are performed
using the PYTHIA signal sample, where the description of the initial state allows
a straightforward definition of the polar angle of the lepton with respect to the quark.
The fully corrected asymmetry values for dileptons at the Born level $\Afb^{\textrm{cor}}$ are shown in \figname~\ref{fig:all:fullunfold}. The magnitude of
the correction is larger than in the previous unfolding step. In addition, the contribution from the PDFs becomes
the dominant systematic uncertainty. Good agreement is observed in general between the measured and
predicted values. The muon channel measurement exhibits a discrepancy
with respect to the PYTHIA prediction for masses above the $Z$~boson mass,
where the measured asymmetry is consistently larger than the prediction.
This effect could not be explained by the analysis procedure and might be a feature of the simulation.

\section{Measurement of \texorpdfstring{\sinsqeff}{sin2thetaWeff}~}
\label{sec:wma}

The extraction of the effective weak mixing angle (\wma{}) from the detector-level asymmetry values
($\Afb^{\textrm{meas}}$) is presented here.

Within the region of interest ($0.218 \leq \sinsqeff \leq 0.236$) 17 MC simulated samples were generated
with varying values of \sinsqeff. The generator used for the templates is PYTHIA, which allows the value of \sinsqeff\ to be tuned without changing $m_{Z}$.
Within the range of the \wma\ variations, the $Z$ boson line shape remains unchanged in the generated samples.
From the generator-level information in the samples, weights are calculated, in bins of $m_{\ell\ell}$ and $\cosThCS$, to transform the
$\Afb^{\textrm{meas}}$~values to the ones expected ($\Afb^{rew}$) for a different value of \wma.
The reweighting technique is validated on simulated samples.
For each channel, the $\Afb^{rew}$~values obtained from the reweighted datasets are compared to
those obtained from the data, using a $\chi^2$ test over the mass range
70--250\,GeV, taking statistical and systematic uncertainties into account.
The mass range has been optimized for the maximum sensitivity and stability of the measurement.
A parabola is fitted to the resulting distributions of $\chi^2$ values for each channel independently.
The minimum of the parabola yields the
\wma\ result. The $\chi^2/ndf$ value at the minimum is
$\wmaChiCC$ for the CC electron channel, $\wmaChiCF$ for the CF electron channel and $\fitquality$ for the muon channel.
The fit results are found to be stable with respect to the invariant mass range over which the
template comparisons are performed, as well as with respect to the \wma\ range over which the $\chi^2$ is minimized.

As discussed in \sectionname~\ref{sec:afbmeas}, the use of a LO generator and a specific implementation of
the real photon emission in the final state does not bias the unfolded $\Afb^{\textrm{meas}}$ values.
In order to assess the impact of these potential sources of systematic effects on the templates used to extract the \wma~value,
additional tests are performed.
Effects related to the modelling of the real photon emission are tested with SHERPA, and are found to be negligible.
The effects of NLO QCD corrections are investigated further in the context of the \sinsqeff~measurement by comparing LO and NLO predictions
of the \Afb\ vs. mass distributions calculated with MCFM~\cite{mcfm}.
Differences are propagated through the extraction of \sinsqeff{} and the resulting variation of the
weak mixing angle is treated as an additional systematic error.
The effects of NLO EWK corrections are found to be small compared to the rest of the uncertainties,
but are accounted for as an additional systematic uncertainty on the final result.

The systematic uncertainties already described for the $\Afb^{\textrm{meas}}$ values are also estimated for the \wma\ measurement.
As the background is small in all channels of the \wma~extraction, a simple, but slightly conservative, approach is used
to obtain its uncertainty.
The \wma~measurement is repeated without the subtraction of the background, and the result is compared
to the baseline measurement, which has the background subtracted.
The uncertainty on \wma\ from the background is taken to be 10\% of the observed difference,
to take into account the uncertainties on the cross sections of the background components known with least
precision~\cite{ZZref}.

\subsection{Impact of PDFs on the \texorpdfstring{\sinsqeff}{sin2thetaWeff}~measurement}

This measurement is sensitive to the PDFs describing the flavour composition of the
initial state, since the \afb\ values depend on the flavour and charge of
the initial partons. In addition, the $u_v$ and $d_v$ valence quark
distributions have an impact on the measurement due to differences in the weak couplings,
while dilution effects introduce a dependence on the sea-to-valence quark ratio within the accessible
Bjorken-{\itshape x} range.

The ATLAS measurements of inclusive $W$ and $Z$ boson production~\cite{WZmeas} are
sensitive to the same effects and indicate that some of the existing PDF sets may not
provide a good description of the data.

For this measurement of \wma\, a LO version of the  ATLAS-epWZ NNLO and NLO
fits was prepared.  In the following, the PDFs extracted from these fits are
called  ATLAS-epWZ12 LO PDFs. These fits include inclusive $W^+$, $W^-$
and $Z$ production data at $\sqrt{s}=7$~TeV~\cite{WZmeas}, together with the combined HERA
data on inclusive neutral- and charged-current interactions from $e^+p$ and $e^-p$ scattering~\cite{HERAPaper}. The
settings for the fit are kept the same as those for the NLO and NNLO fits as much
as possible.
The main differences between the LO fit and the higher-order fits are that the gluon PDF distribution
has a simple parameterization which requires it to
be positive definite at all scales, and that the value of the strong coupling constant $\alpha_{\mathrm{s}}(m_Z)$
is set to be higher
than the one for the NLO or NNLO fits. A value of $\alpha_{\mathrm{s}}(m_Z)$ of 0.130 is chosen with a scanning procedure,
such that it yields the best level of agreement ($\chi^2$) between data and the fit result.
This value is consistent with that used by other LO PDF sets.
These PDFs are available in LHgrid formats~\cite{LHAPdf}.

\subsection{Results for \texorpdfstring{\sinsqeff}{sin2thetaWeff}~}

The measured values of the weak mixing angle obtained with ATLAS-epWZ12 LO PDF are
shown in \tabname~\ref{tab:ATLAS_resultComparison}.
Of the three channels, the CF electron channel has the lowest statistical uncertainty, despite
having the fewest selected events, as detailed in~\sectionname~\ref{sec:eventselection}.
This is because the $\Afb^{\textrm{meas}}$~values in this channel are less affected by dilution due to the
larger average rapidity of the dilepton system compared to the other two channels.
Details of the main sources of systematic uncertainty on the result in each channel are given in
\tabname~\ref{tab:wmasyst}.

The \wma~measurements from all three channels are combined. Given the total covariance matrix $C$, the weights assigned
to the three measurements for the combination
are calculated using $w_i=\sum _k {C}^{-1}_{ik}/\sum _{jk} C^{-1}_{jk}$, with the indices $i$, $j$ and $k$
representing the three measurements (CC electron, CF electron, muon).
The total error is given by $\sigma ^2=(W^{\mathrm{T}} C^{-1})^{-1}W$, with $W^{\mathrm{T}}=(1,1,1)$.
Uncertainties due to energy/momentum scale and resolution
 are treated as completely uncorrelated across channels. Since the energy scale and resolution for the
CF electron channel is dominated by the forward calorimeter performance,
this is a good approximation for the correlation of the CC and CF electron
channels. The systematic uncertainties due to the MC statistical uncertainty are also
  treated as fully uncorrelated. Theory-related systematic uncertainties, such as those associated with PDFs and higher-order EWK and QCD corrections,
 are treated as fully correlated across channels.
 All the
 remaining uncertainties are treated as fully correlated across the channels to which they are applicable.
The central value of the combination is found to be stable when varying the magnitudes
of the main correlated uncertainties. Following this procedure, the combined (CC+CF) electron result, as well as the
electron--muon combination, is shown in \tabname~\ref{tab:ATLAS_resultComparison}. The systematic uncertainty on the
combined result is dominated by the PDF uncertainty $(\pm 0.0009)$, which is calculated using the variations provided
in the ATLAS-epWZ12 LO PDF eigenvalue set,
similar to the calculation described in section \ref{sec:afbmeas}.
The contributions to the total systematic uncertainties on the results are included in \tabname~\ref{tab:wmasyst}.

The results obtained with the ATLAS-epWZ12 LO PDF set have been compared to those obtained using other
leading order PDF sets. The PDF sets ATLAS-epWZ12 LO and HERAPDF1.5LO~\cite{HERAPDF} yield very similar results,
the ATLAS-epWZ12 LO result being larger by $1 \times 10^{-4}$ (with a PDF uncertainty of $9 \times 10^{-4}$).
The MSTW2008LO set produces a downward shift in the resulting \wma\ which is significant
($\Delta\wma=-2\times 10^{-3}$). However, MSTW sets are known to give a poor description
of the ATLAS $W$ and $Z$ data~\cite{WZmeas}.
The CT10 PDF set (which is NLO) was also tested, and yields a \wma\ value which is smaller
than,
but compatible with, the results obtained with the ATLAS-epWZ12 LO PDF ($\Delta\wma= -8\times 10^{-4}$).
The PDF analysis of the ATLAS data also suggests an increased
strange-quark sea density ($\bar{s}/\bar{d} \sim 1$) compared to other PDF sets~\cite{epWZpdf}.
In order to probe the sensitivity of this measurement to the enhanced strange-quark density, a
dedicated PDF set was prepared with a suppressed strange sea corresponding to
uncertainty, which indicates a low sensitivity of the measurement to this effect.
This is due to the fact that the sea composition only affects the measurement
through the dilution, which, to first approximation, does not change the position of the minimum
of the $\chi^2$ in the template fits.

As a cross-check of the measured weak mixing angle and the unfolding procedure, \wma\ is
extracted from the unfolded particle-level asymmetries.
Similarly to the procedure described in the beginning of section \ref{sec:wma},
samples with various values of \wma, generated using MCFM v6.8~\cite{mcfm}  at
LO in perturbative QCD and interfaced to APPLGRID v1.4.69~\cite{applgrid} using the ATLAS-epWZ12 LO PDF set,
are compared to the measured particle-level asymmetries using a $\chi^2$ test.
The correlations between the systematic uncertainties are included in the $\chi^2$ calculation. A parabolic fit to the $\chi^2$
distribution yields a value of \wma\ in good agreement with the detector-level extraction.

\begin{figure}[ht]
\centering
	\includegraphics[width=0.7\textwidth]{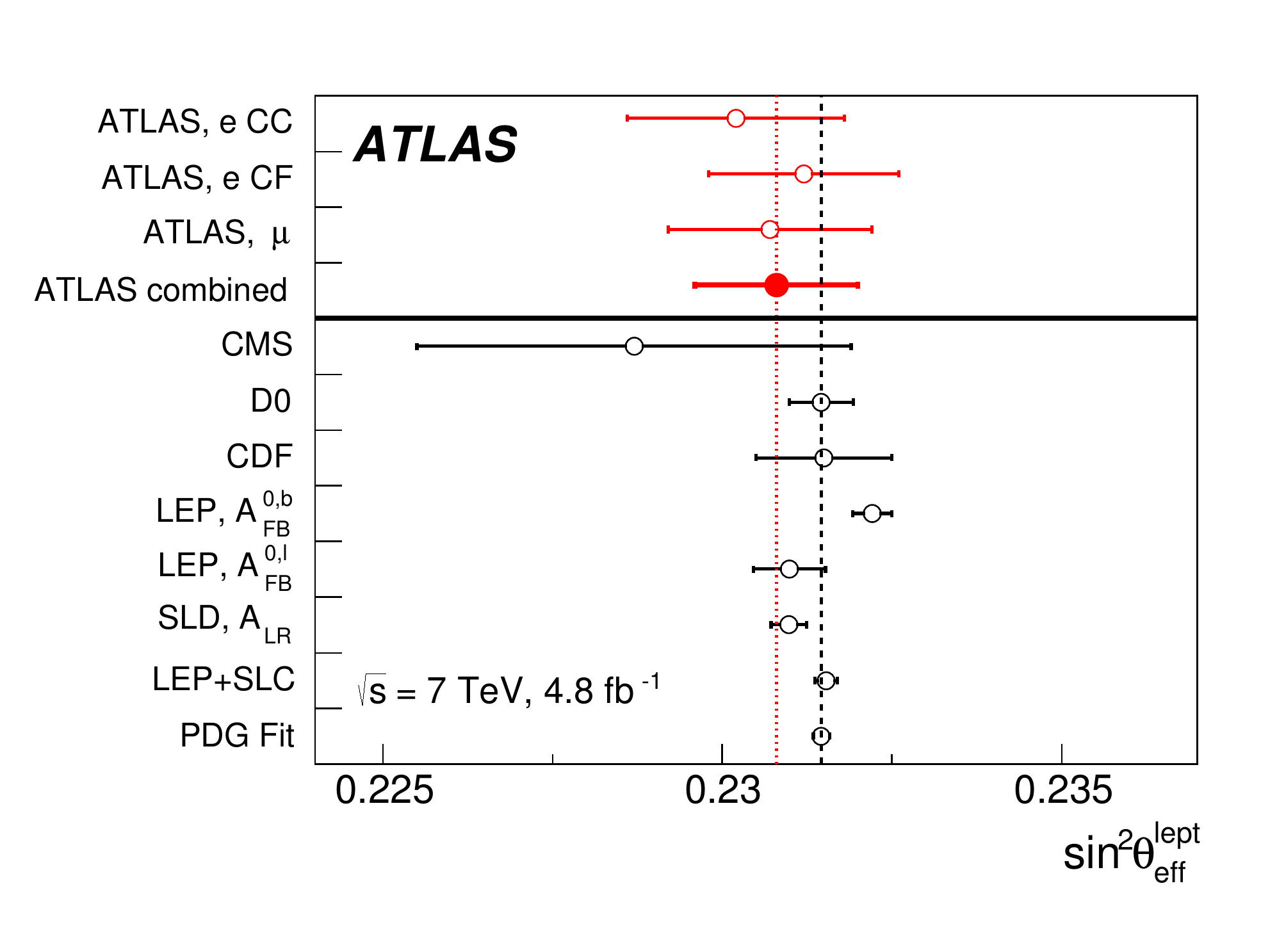}
	\caption{
	Comparison of the results of this analysis with other published results for $\wma$.
	This includes the most precise measurements from LEP and SLC, and the leptonic
	\wma~measurements from the hadron collider experiments CMS~\cite{CMSNote},
	D0~\cite{D0sin2ThW}, and CDF~\cite{CDFsin2ThWMu}. Also shown are the values of \wma~from
	the LEP+SLC global combination~\cite{EWKmeasurementsZ} (which includes all \wma\
	measurements performed at the two colliders) and from the PDG global fit~\cite{pdg:effWeakAngle}.
	The vertical dotted line
	shows the central value of the ATLAS combined measurement reported here, while the vertical
	dashed line represents that of the current PDG global fit~\cite{pdg:effWeakAngle}. }
	\label{fig:result_comparison}
\end{figure}

There are several other measurements of \wma~at the $m_{Z}$ scale.
The results from this analysis, as well as the results from the other collider experiments,
are summarized in \tabname~\ref{tab:afb_resultComparison} and shown in~\figname~\ref{fig:result_comparison}.
These include measurements from LEP, SLC, the Tevatron, and the LHC, as well as the results of the PDG global fit~\cite{pdg:effWeakAngle}.
Differences with respect to the combined LEP/SLC measurements are also displayed.
The combined result of this analysis agrees within $0.1\,\sigma$ with the
most precise leptonic asymmetry measurement, and within $1.2\,\sigma$ of the \wma~value extracted from the LEP $\Afb^{0,b}$ measurement.
The combined result is in agreement with the PDG global fit at the level of $0.6\,\sigma$.

\begin{table}[htbp]
\begin{center}
	\begin{tabular}{l|c}
	\hline \hline
	 & $\wma$ \\ \hline \hline
	CC electron 		& $\wmaElVATLAS  \pm \wmaElStat  \mathrm{(stat.)} \pm \wmaElSysNoPDFATLAS \mathrm{(syst.)}  \pm \wmaElSysPDFATLAS \mathrm{(PDF)} = \wmaElVATLAS  \pm \wmaElErrorATLAS$ \\
	CF electron 		& $\wmaElVCFATLAS \pm \wmaElStatCF \mathrm{(stat.)} \pm \wmaElSysNoPDFCFATLAS \mathrm{(syst.)} \pm \wmaElSysPDFCFATLAS \mathrm{(PDF)} = \wmaElVCFATLAS  \pm \wmaElErrorCFATLAS$ \\
	Muon			& $\wmaMuVATLAS \pm \wmaMuStat \mathrm{(stat.)} \pm \wmaMuSysNoPDFATLAS \mathrm{(syst.)} \pm \wmaMuSysPDFATLAS \mathrm{(PDF)} = \wmaMuVATLAS  \pm \wmaMuErrorATLAS$ \\
	El. combined 	& $\wmaCombEATLAS  \pm \wmaCombEStat  \mathrm{(stat.)} \pm \wmaCombESysNoPDFATLAS \mathrm{(syst.)} \pm \wmaCombESysPDFATLAS \mathrm{(PDF)} = \wmaCombEATLAS  \pm \wmaCombEErrorATLAS$ \\ \hline
	Combined 		& $\wmaCombATLAS  \pm \wmaCombStat  \mathrm{(stat.)} \pm \wmaCombSysNoPDFATLAS \mathrm{(syst.)} \pm \wmaCombSysPDFATLAS \mathrm{(PDF)} = \wmaCombATLAS  \pm \wmaCombErrorATLAS$ \\
	\hline\hline
	\end{tabular}
	\caption{The $\wma$~measurement results in each of the three studied channels:
	electron central-central, electron central-forward and muon.
	Results of the statistical combination of both electron channels and all three channels
	are shown as well.}
	\label{tab:ATLAS_resultComparison}
\end{center}
\end{table}

\begin{table}[htbp]
\begin{center}
		\begin{tabular}{l|c|c|c|c}
	\hline\hline
		 		& CC electrons 	& CF electrons 		& Muons 		& Combined  \\
	 Uncertainty source &  [$10^{-4}$]	&  [$10^{-4}$]		&   [$10^{-4}$]		&   [$10^{-4}$] \\
	\hline\hline
	PDF 		&		 10	&	10		&	9		& 9 \\
	MC statistics &			 5	&	2		&	5		& 2 \\
	Electron energy scale &		 4	&	6		&	--		& 3 \\
	Electron energy resolution &	 4	&	5		&	--		& 2 \\
	Muon energy scale 	&	 --	&	--		&	5		& 2\\
	Higher-order corrections &	 3	&	1		&	3		& 2 \\
	Other sources &			 1	&	1		&	2		& 2\\
	\hline
	\end{tabular}
	\caption{Contributions to the systematic uncertainties on the \wma\ values extracted from the three analysis channels and on the combined result.
	Null entries (denoted by ``--'') correspond to uncertainties that do not apply to a specific channel.
     	Higher-order corrections include NLO QCD and NLO EWK contributions. Other sources include the effect of pileup, background uncertainties, lepton
     	trigger/reconstruction/identification efficiency uncertainties, muon momentum resolution and effects of detector misalignment.
	}
	\label{tab:wmasyst}
\end{center}
\end{table}

\begin{table}[htbp]
\begin{center}
		\begin{tabular}{l|c|S[table-format=3.2]|S[table-format=3.2]}
	\hline\hline
		 &   & $\Delta / \sigma$  & $\Delta / \sigma$ \\
	       	&    $\wma$   &   \multicolumn{1}{c|}{(wrt LEP+SLC) }   & \multicolumn{1}{c}{ (wrt ATLAS)} \\
	\hline\hline
	CMS \cite{CMSNote} & $0.2287 \pm 0.0032$  & -0.9& -0.6 \\
	D0 \cite{D0sin2ThW} 	& $0.23146 \pm 0.00047$  &  -0.1 & 0.5 \\
	CDF \cite{CDFsin2ThWMu}	& $0.2315\pm0.0010$ & -0.03  & 0.4 \\
	LEP, $\Afb^{0,b}$ \cite{EWKmeasurementsZ}& $0.23221 \pm 0.00029$ & \multicolumn{1}{c|}{ -- } & 1.2\\
	LEP, $\Afb^{0,l}$ \cite{EWKmeasurementsZ}& $0.23099 \pm 0.00053$ & \multicolumn{1}{c|}{ -- } & -0.1\\
	SLC, $A_{\mathrm LR}$ \cite{EWKmeasurementsZ}& $0.23098 \pm 0.00026$ & \multicolumn{1}{c|}{ -- } & -0.1\\
	LEP+SLC \cite{EWKmeasurementsZ}& $0.23153 \pm 0.00016$ & \multicolumn{1}{c|}{ -- } & 0.6\\
	\hline\hline
	PDG global fit \cite{pdg:effWeakAngle} & $0.23146 \pm 0.00012$ & -0.4 & 0.6\\
	\hline\hline
	ATLAS & $\wmaCombATLAS \pm \wmaCombErrorATLAS$ & -0.6 &\multicolumn{1}{c}{ -- } \\

	\end{tabular}
	\caption{
	Comparison of the results of this analysis with other published results for $\wma$. The comparison includes the most precise measurements from LEP and SLC, and
	the results from the leptonic
	\wma~measurements from the hadron collider experiments CMS, D0, and CDF. Also shown are the
	values of \wma~from the LEP+SLC global
	combination (which includes all \wma\ measurements performed at the two colliders) and from the
	PDG global fit. Each $\Delta / \sigma$ column shows the difference between the result and the
	quoted reference value, divided by the quadratic sum of the associated uncertainties.}
	\label{tab:afb_resultComparison}
\end{center}
\end{table}

\subsection{Determination of \texorpdfstring{\AmuPar}{AmuPar}~}
\label{sec:Amu}

As described in \sectionname~\ref{sec:intro}, the forward-backward asymmetry around the $Z$ pole for the muon
channel, $A^{0,\mu}_{\mathrm{FB}}$, can be expressed in terms of the muon and quark asymmetry parameters $A_\mu$ and
$A_q$, c.f. \eqname~\ref{eq:ai} and ref.~\cite{EWKmeasurementsZ}. Both are functions of the vector and
axial-vector couplings of the quark/muon,

\begin{equation}\label{eq:aq}
 A_{q/\mu} =  \frac{ 2 g_{\mathrm{V}}^{q/\mu} g_{\mathrm{A}}^{q/\mu}} {(g_{{\mathrm{V}}}^{q/\mu})^2 + (g_{\mathrm{A}}^{q/\mu})^2} =
\frac{ 2 g_{\mathrm{V}}^{q/\mu} / g_{\mathrm{A}}^{q/\mu} } { 1+
(g_{\mathrm{V}}^{q/\mu}/g_{\mathrm{A}}^{q/\mu})^2}.
\end{equation}
The asymmetry parameters are related to the flavour-dependent
weak mixing angle $\wmaqmu$ by

\begin{equation}\label{eq:gv}
g_{\mathrm{V}}^{q/\mu}/g_{\mathrm{A}}^{q/\mu} = 1 - 4 |Q_{q/\mu}| \wmaqmu\ ,
\end{equation}

\noindent where $Q_{q/\mu}$ is the quark ($q$) or muon ($\mu$) charge.
The measurement of the
forward-backward asymmetry can be interpreted as a determination of $A_\mu$ when assuming
$\wmaq$ = $\wmamu$ = $\wma$, which
is valid within an uncertainty of $1.5 \times 10^{-4}$~\cite{EWKmeasurementsZ}.
Given the value of $\wma$,
$A_{\mu}$ is small compared to $A_{q}$
and the uncertainty on $\wmamu$ ($\sim 1.5 \times 10^{-3}$) plays a more important role in
\eqname~\ref{eq:ai} than the error introduced by the assumption about $\wmaq$ mentioned above.
With the present precision, it is therefore possible to use the value of \wmamu\ obtained from the muon
asymmetry measurement to derive

\begin{equation}\label{eq:amu}
A_\mu = \frac{2(1-4\wma)} {  1 + (1 - 4 \sinsqeff)^{2}}.
\end{equation}
The uncertainties on $A_\mu$ are
propagated from the uncertainties on \wma.
The result is

 \[\AmuPar= \AmuParATLAS \pm \AmuParErrorStatATLAS \mathrm{(stat.)} \pm
 \AmuParErrorSysATLAS \mathrm{(syst.)} \pm \AmuParErrorPDFATLAS \mathrm{(PDF)}= \AmuParATLAS \pm \AmuParErrorTotATLAS \mathrm{(tot.)},\]
which is of similar precision and in good agreement with the measurement from $e^+e^-$
collisions of $0.142\pm 0.015$ \cite{EWKmeasurementsZ}.
It is worth stressing, however, that the determination of \AmuPar\ in the LEP/SLD results
is based entirely on asymmetry measurements in the different lepton
final states without any assumptions on other $A_f$, whereas the determination of \AmuPar\
presented here uses the Standard Model prediction of $A_q$.

\section{Conclusions}
The forward-backward asymmetry in electron and muon pairs from $\zgamma$ decays
is measured using
the 7\,TeV $pp$ LHC collision data recorded with the ATLAS detector in 2011 corresponding to an integrated luminosity
of 4.8 fb$^{-1}$.
The data are analysed over a range of dilepton invariant masses from 66~GeV to 1000~GeV in the central-central electron and muon channels, and up to 250 GeV
in the central--forward electron channel. The latter includes events where one electron is reconstructed in the forward pseudorapidity range ($2.5<|\eta|<4.9$).

The forward-backward asymmetry is measured separately for the three channels as a
function of the dilepton invariant mass and unfolded for detector effects and final-state radiation.
Additionally, a leading-order interpretation which accounts for the effects of dilution and full detector acceptance is presented.
The resulting \Afb~values are found to be in agreement with the corresponding Standard Model predictions.

The detector level asymmetry values are used to extract the value of the leptonic effective weak mixing angle, \sinsqeff, separately
for the three data samples using a $\chi^2$ minimization method. The results are in good agreement with each other and with measurements
at $e^+e^-$ colliders, at the Tevatron and by CMS at the LHC.

Results from the electron and muon final states are combined, yielding

\[ \sinsqeff= \wmaCombATLAS  \pm \wmaCombStat  \mathrm{(stat.)} \pm \wmaCombSysNoPDFATLAS \mathrm{(syst.)} \pm \wmaCombSysPDFATLAS \mathrm{(PDF)}  =  \wmaCombATLAS  \pm \wmaCombErrorATLAS \mathrm{(tot.)}.\]
The dominant uncertainty comes from knowledge of the PDFs.

The result from the muon channel, when converted to the asymmetry parameter \AmuPar, yields

\[ \AmuPar = \AmuParATLAS \pm \AmuParErrorStatATLAS \mathrm{(stat.)} \pm \AmuParErrorSysATLAS \mathrm{(syst.)} \pm \AmuParErrorPDFATLAS \mathrm{(PDF)}
 = \AmuParATLAS \pm \AmuParErrorTotATLAS \mathrm{(tot.)},\]

\noindent which is in good agreement with the best previous measurements.

\FloatBarrier

\section*{Acknowledgements}

We thank CERN for the very successful operation of the LHC, as well as the
support staff from our institutions without whom ATLAS could not be
operated efficiently.

We acknowledge the support of ANPCyT, Argentina; YerPhI, Armenia; ARC, Australia; BMWFW and FWF, Austria; ANAS, Azerbaijan; SSTC, Belarus; CNPq and FAPESP, Brazil; NSERC, NRC and CFI, Canada; CERN; CONICYT, Chile; CAS, MOST and NSFC, China; COLCIENCIAS, Colombia; MSMT CR, MPO CR and VSC CR, Czech Republic; DNRF, DNSRC and Lundbeck Foundation, Denmark; IN2P3-CNRS, CEA-DSM/IRFU, France; GNSF, Georgia; BMBF, HGF, and MPG, Germany; GSRT, Greece; RGC, Hong Kong SAR, China; ISF, I-CORE and Benoziyo Center, Israel; INFN, Italy; MEXT and JSPS, Japan; CNRST, Morocco; FOM and NWO, Netherlands; RCN, Norway; MNiSW and NCN, Poland; FCT, Portugal; MNE/IFA, Romania; MES of Russia and NRC KI, Russian Federation; JINR; MESTD, Serbia; MSSR, Slovakia; ARRS and MIZ\v{S}, Slovenia; DST/NRF, South Africa; MINECO, Spain; SRC and Wallenberg Foundation, Sweden; SERI, SNSF and Cantons of Bern and Geneva, Switzerland; MOST, Taiwan; TAEK, Turkey; STFC, United Kingdom; DOE and NSF, United States of America. In addition, individual groups and members have received support from BCKDF, the Canada Council, CANARIE, CRC, Compute Canada, FQRNT, and the Ontario Innovation Trust, Canada; EPLANET, ERC, FP7, Horizon 2020 and Marie Skłodowska-Curie Actions, European Union; Investissements d'Avenir Labex and Idex, ANR, Region Auvergne and Fondation Partager le Savoir, France; DFG and AvH Foundation, Germany; Herakleitos, Thales and Aristeia programmes co-financed by EU-ESF and the Greek NSRF; BSF, GIF and Minerva, Israel; BRF, Norway; the Royal Society and Leverhulme Trust, United Kingdom.

The crucial computing support from all WLCG partners is acknowledged
gratefully, in particular from CERN and the ATLAS Tier-1 facilities at
TRIUMF (Canada), NDGF (Denmark, Norway, Sweden), CC-IN2P3 (France),
KIT/GridKA (Germany), INFN-CNAF (Italy), NL-T1 (Netherlands), PIC (Spain),
ASGC (Taiwan), RAL (UK) and BNL (USA) and in the Tier-2 facilities
worldwide.

\clearpage\addcontentsline{toc}{section}{\numberline{}{Bibliography}}
\printbibliography
\clearpage

\FloatBarrier
\newpage
\begin{flushleft}
{\Large The ATLAS Collaboration}

\bigskip

G.~Aad$^{\rm 84}$,
B.~Abbott$^{\rm 112}$,
J.~Abdallah$^{\rm 152}$,
S.~Abdel~Khalek$^{\rm 116}$,
O.~Abdinov$^{\rm 11}$,
R.~Aben$^{\rm 106}$,
B.~Abi$^{\rm 113}$,
M.~Abolins$^{\rm 89}$,
O.S.~AbouZeid$^{\rm 159}$,
H.~Abramowicz$^{\rm 154}$
% ...
\bigskip
\\
% ...
$^{11}$ Institute of Physics, Azerbaijan Academy of Sciences, Baku, Azerbaijan\\
% ...
$^{84}$ CPPM, Aix-Marseille Universit{\'e} and CNRS/IN2P3, Marseille, France\\
% ...
$^{89}$ Department of Physics and Astronomy, Michigan State University, East Lansing MI, United States of America\\
% ...
$^{106}$ Nikhef National Institute for Subatomic Physics and University of Amsterdam, Amsterdam, Netherlands\\
% ...
$^{112}$ Homer L. Dodge Department of Physics and Astronomy, University of Oklahoma, Norman OK, United States of America\\
$^{113}$ Department of Physics, Oklahoma State University, Stillwater OK, United States of America\\
% ...
$^{116}$ LAL, Universit{\'e} Paris-Sud and CNRS/IN2P3, Orsay, France\\
% ...
$^{152}$ Institute of Physics, Academia Sinica, Taipei, Taiwan\\
% ...
$^{154}$ Raymond and Beverly Sackler School of Physics and Astronomy, Tel Aviv University, Tel Aviv, Israel\\
% ...
$^{159}$ Department of Physics, University of Toronto, Toronto ON, Canada\\
% ...
\end{flushleft}


\end{document}